\documentclass[smallextended]{svjour3}       
\smartqed  
\usepackage{graphicx}
\usepackage{xcolor}
	\usepackage{pgfplots}
\pgfplotsset{compat=1.13} 
\usepackage{pgfplotstable}
\pgfplotscreateplotcyclelist{my color list}{%
	solid, color=webblue, every mark/.append style={solid, fill=webblue}, mark=*\\%
	densely dashdotted, color=webred, every mark/.append style={solid, fill=webred},mark=diamond*\\%
	densely dotted, color=webgreen, every mark/.append style={solid, fill=webgreen}, mark=triangle*\\%
	loosely dashed, color=webbrown, every mark/.append style={solid, fill=webbrown},mark=square,fill=webgreen!20\\%
	densely dotted, every mark/.append style={solid, fill=gray}, mark=otimes*\\%
	dashed, every mark/.append style={solid, fill=gray},mark=diamond*\\%
	densely dashed, every mark/.append style={solid, fill=gray},mark=square*\\%
	dashdotted, every mark/.append style={solid, fill=gray},mark=otimes*\\%
	dashdotdotted, every mark/.append style={solid},mark=star\\%
}
\definecolor{webgreen}{rgb}{0,.5,0}
\definecolor{webbrown}{rgb}{.6,0,0}
\definecolor{webyellow}{rgb}{0.98,0.92,0.73}
\definecolor{webgray}{rgb}{.753,.753,.753}
\definecolor{webblue}{rgb}{0,0,.8}
\definecolor{webgreen}{rgb}{0, 0.5, 0} 
\definecolor{webred}{rgb}{0.5, 0, 0}   

\usepackage[toc,acronym]{glossaries}
\usepackage{adjustbox}
\usepackage{tikz}
\usetikzlibrary{arrows,shadows,backgrounds,calc,decorations.pathreplacing,decorations.pathmorphing,positioning, 
	automata}
\newcommand*{\boxcolor}{orange}
\makeatletter
\renewcommand{\boxed}[1]{\textcolor{\boxcolor}{%
		\tikz[baseline={([yshift=-1ex]current bounding box.center)}] \node [rectangle, minimum width=1ex,rounded corners,draw] {\normalcolor\m@th$\displaystyle#1$};}}
\makeatother

\usepackage{pgfplots}
\pgfplotsset{compat=1.13} 
\usepackage{pgfplotstable}
\makeglossaries

\newacronym{AI}{AI}{Artificial Intelligence}
\newacronym{ANN}{ANN}{Artificial Neural Network}
\newacronym{FPGA}{FPGA}{Field Programmable Gate Array}  
\newacronym{CPU}{CPU}{Central Processing Unit}
\newacronym{GPU}{GPU}{Graphic Processing Unit}
\newacronym{HW}{HW}{hardware}
\newacronym{ISA}{ISA}{Instruction Set Architecture}
\newacronym{I/O}{I/O}{Input/Output}
\newacronym{MC}{MC}{Multi-Core and/or Many-Core}
\newacronym{MLP}{MLP}{Memory Level Parallelism}
\newacronym{OoO}{OoO}{Out-of-Order}
\newacronym{OS}{OS}{operating system}
\newacronym{PO}{PO}{Propagation Overhead}
\newacronym{PD}{PD}{Propagation  Delay}
\newacronym{PU}{PU}{Processing Unit}
\newacronym{SPA}{SPA}{Single Processor Approach}
\newacronym{SW}{SW}{software}
\newacronym{HPL}{HPL}{High Performance Linpack}
\newacronym{HPCG}{HPCG}{High Performance Conjugate Gradients}

\begin{document}

\title{Finally, how many efficiencies the supercomputers have?\thanks{Project no. 136496 has been implemented with the support provided from the National Research, Development and Innovation Fund of Hungary, financed under the K funding scheme.}
}
\subtitle{And, what do they measure?\thanks{A follow-up version of paper The Journal of Supercomputing volume 76, pages 9430–9455 (2020) }}

\titlerunning{The many efficiencies of supercomputers}  

\author{J\'anos V\'egh
}

\institute{J. V\'egh \at
              Kalim\'anos BT, Hungary \\
              \email{Vegh.Janos@gmail.com}           
}

\date{Received: date / Accepted: date}

\maketitle

\begin{abstract}
Using a vast number of processing elements in computing systems led to unexpected phenomena, such as different efficiencies of the same system for various tasks,
which cannot be explained in the frame of classical
computing paradigm.
The simple, non-technical (but considering the temporal behavior of the components) model,
introduced here, enables us to set up a frame and formalism needed to explain those unexpected experiences around
supercomputing.
Introducing temporal behavior into computer science also explains why only the extreme-scale computing
enabled us to discover the experienced limitations.
The paper shows that the degradation of efficiency of parallelized sequential systems
is a natural consequence of the classical computing paradigm instead of being an engineering imperfectness.
The workload that supercomputers run is responsible for wasting energy and limiting the size and type of tasks.
Case studies provide insight into how different contributions compete for dominating
the resulting payload performance of a computing system
and how enhancing the interconnection technology made computing+communication
dominate in defining the efficiency of supercomputers.
Our model also enables us to derive predictions
about supercomputer performance limitations for the near future,
as well as it provides hints for enhancing supercomputer components. 
Phenomena experienced in large-scale computing show interesting parallels
with phenomena experienced in science more than a century ago,
and through their studying, developed a modern science.

\end{abstract}

\section{Introduction}
\label{intro}

Given that the dynamic growth of single-processor performance has stalled about two decades ago~\cite{ComputingPerformanceBook:2011}
and the computing demand grows more speedily than the 
computing capacity~\cite{BrainMasterPlan:2022},
the only way to achieve the required high computing performance
remained to parallelize the work of a vast number of
sequentially working single processors.
However, as was very early predicted~\cite{AmdahlSingleProcessor67}, and decades later
experimentally confirmed~\cite{ScalingParallel:1993}, the
\textit{scaling} of parallelized computing is not linear.
Even, "\textit{there comes a point when using more processors \dots actually increases the execution time rather than reducing it}"~\cite{ScalingParallel:1993}. 
Parallelized sequential processing has different rules of game~\cite{ScalingParallel:1993},~\textbf{\cite{VeghModernParadigm:2019}}: its performance gain ("speedup") has its inherent bounds~\textbf{\cite{VeghPerformanceWall:2019}}.

Akin to as laws of science limit the performance
of single-thread processors~\cite{LimitsOfLimits2014},
the commonly used computing paradigm (through its technical implementation) limits the payload performance of supercomputers~\textbf{\cite{VeghModernParadigm:2019}}.
On the one side, experts expected performance\footnote{There are doubts about the definition of exaFLOPS, whether it means nominal performance $R_{Peak}$ or payload performance $R_{Max}$; measured
	by which benchmark (or how it depends on the workload the computer runs); using which operand length. Here the term is used as $R_{Max}^{HPL-64}$. Several other benchmarks results (not related to floating computation)  have been published to produce higher figures.} to achieve the magic 1~Eflops
around year 2020, Figure~1 in~\cite{ChinaExascale:2018}\footnote{A special issue https://link.springer.com/journal/11714/19/10}.
"\textit{The performance increase of the No. 1 systems slowed down around 2013, and it was the same for the sum performance}"~\cite{ChinaExascale:2018}, but the authors extrapolated linearly, expecting that the development continues and shall achieve "zettascale computing" (i.e., $10^4$-fold more than the present performance) in just more than a decade. On the other side, it has recently been admitted that linearity is "\textit{A trend that can't go on ad infinitum.}" Furthermore, that it "\textit{can be seen in our current situation where the historical ten-year cadence between the attainment of megaflops, teraflops, and petaflops has not been the case for exaflops}"\cite{ExascaleGrandfatherHPC:2019}.
Officially, the TOP500~\cite{TOP500} evaluation sounds (as of 2020) that "\textit{the top of the list remains largely unchanged}" and "\textit{the full list recorded the smallest number of new entries since the project began in 1993}".
The 2021 list added: "\textit{Still waiting for Exascale}".

The expectations against supercomputers are excessive. For example, the name of the company PEZY\footnote{https://en.wikipedia.org/wiki/PEZY\_Computing: The name PEZY is an acronym derived from the greek derived Metric prefixes Peta, Eta, Zetta, Yotta} witnesses that a billion times increase in payload performance is expected.
It looks like that in the feasibility studies on supercomputing using
parallelized sequential systems, an analysis
whether building computers of such size is feasible (and reasonable) remained out of sight either in USA~\cite{NSA_DOE_HPC_Report_2016,Scienceexascale:2018} or in EU~\cite{EUActionPlan:2016}
or in Japan~\cite{JapanExascale:2018} or in China~\cite{ChinaExascale:2018}.
Even in the most prestigious journals, the "gold rush" is going on~\cite{StrechingSupercomputers:2017,Scienceexascale:2018}.
In addition to the previously existing "\textit{two different efficiencies of supercomputers}"~\cite{DifferentBenchmarks:2017}
other efficiency/performance values appeared\footnote{https://blogs.nvidia.com/blog/2019/06/17/hpc-ai-performance-record-summit/\\ https://www.olcf.ornl.gov/2018/06/08/genomics-code-exceeds-exaops-on-summit-supercomputer/} (of course with higher numeric figures), and we can easily derive several more efficiencies.

Although severe counter-arguments were also published, mainly based on the power consumption of single processors and large computing centers~\cite{WhyNotExascale:2014}, the moon-shot of limitless parallelized processing performance is followed. 
The probable source of the idea is the "weak scaling"~\cite{Gustafson:1988,ScalingParallel:1993}\footnote{The related work and speedup deserved the Gordon Bell Prize}. However, it is based simply on a misinterpretation~\cite{UsesAbusesAmdahl:2001,AmdalVsGustafson96} of terms in Amdahl's law~\textbf{\cite{VeghScalingANN:2021}}\footnote{As explicitly suspected in~\cite{AmdalVsGustafson96}: \textit{Gustafson's formulation gives an illusion that as if N can increase indefinitely}.}. 
In reality, \textit{Amdahl's Law (in its original spirit) is valid for all parallelized sequential activities, including computing-unrelated ones, and it is the governing law of distributed (including super-) computing}.

Demonstrative failures of some 
systems (such as supercomputers Gyoukou and Aurora'18\footnote{It was also learned that \textit{specific processor design is needed for exascale}
	As part of the announcement, the development line \textit{Knights Hill}~\cite{IntelDumpsXeonPhi:2017} was canceled and instead, be replaced by a "new platform and new micro-architecture specifically designed for exascale"}, and brain simulator SpiNNaker\footnote{Despite its failure, SpiNNaker2 is also under construction~\cite{SpiNNaker2:2018}}) are already known, and we expect many more to follow: such as Aurora'21~\cite{DOEAurora:2017},
the mystic China supercomputers\footnote{https://www.scmp.com/tech/policy/article/3015997/china-has-decided-not-fan-flames-super-computing-rivalry-amid-us\\
	https://www.nextplatform.com/2021/11/15/why-did-china-keep-its-exascale-supercomputers-quiet/
}. Fugaku, although it considerably enhanced the efficacy of computing,
mainly due to the clever placing and use mode of its on-chip memory, also stalled at about 40\% of its planned capacity~\cite{DongarraFugakuSystem:2020} 
and could increase its payload performance only marginally in a year.
Perlmutter produced a modest 10\% payload performance enhancement in its Phase 2.
No sign of life was seen from US exascale supercomputer Aurora.

Similar is the case with exascale applications, such as brain simulation.
Exaggerated news about simulating the brain of some animals
or a large percentage of the human brain appeared.
The reality is that the many-thread implementation of the brain simulator
can fill a tremendous amount of memory with data of billions of artificial neurons~\cite{SpikingPetascale2014}, a purpose-built (mainly \gls{HW}) brain simulator can be designed to simulate one billion neurons~\cite{SpiNNaker:2013}.
However, in practice, they both can simulate only about 80 thousand neurons~\cite{NeuralNetworkPerformance:2018}, mainly because of
"the quantal nature of the computing time"~\textbf{\cite{VeghBrainAmdahl:2019}}.
"More Is Different"~\cite{MoreIsDifferent1972}.

In June 2022, Frontier achieved the magic barrier. The major contribution was a seamless integration by sharing $L_1$ and $L_2$ data of \gls{CPU} with \gls{GPU}. Although this kind of accelerated processors eliminated the performance-limiting effect of the
former implementations (particularly at high number of \gls{PU}s) and caused a theoretically unpredictable performance jump in supercomputers' performance, it needed no changes in their theoretical discussion.

The confusion keeps growing.
The paper attempts to clarify the terms by
scrutinizing the basic notions, contributions, and measurement methods.
In section~\ref{sec:TheModel} a by intention enormously simplified
non-technical model, based on the temporal behavior of a physical implementation of computing~\textbf{\cite{VeghTemporal:2020}}, is presented.
The notations for Amdahl's Law, which form the basis of the present paper,
are introduced in section~\ref{sec:AmdahlsLaw}. We show
that \textit{the degradation of efficiency of parallelized sequential systems} \textit{is a natural consequence of the computing paradigm} instead of being
an engineering imperfectness (in the sense that it could be fixed later).
Furthermore, its consequence is that the parallelized 
sequential computing systems by their very nature 
have an upper \textit{payload performance} (or more precisely, a payload \textit{performance gain}) bound.
Different contributions arise from the sequential portion of the task (and through this, they degrade its parallelized performance),
as detailed in section~\ref{sec:Contributions}.
We validate the established model in section~\ref{sec:accuracy}.

Given that the race to produce computing systems, having components and systems with higher performance numbers, is going on,
in section \ref{sec:zettaflops}, the expectable results of developments in the near future are predicted. The section introduces some further performance merits and, through interpreting them, concludes that 
increasing the size of supercomputers further and making 
expensive enhancements in their technologies, only increase 
their \textit{non-payload performance}.
In section~\ref{sec:Analogies} we discuss that
under extreme conditions, technical objects of computing show up in a series of
behavioral features (for more details see~\textbf{\cite{VeghModernParadigm:2019}}), similar to those
of natural objects. 

\section{A conceptual model of parallelized sequential operation}

The performance measurements are simple time measurements\footnote{Sometimes also secondary merits, such as GFlops/Watt or GFlops/USD are also derived} (although they need careful handling and proper interpretation, see good textbooks such as~\cite{RISCVarchitecture:2017}): a standardized set of machine instructions is executed
(a large number of times) and the known number of \textit{relevant} operations are divided by the measurement time;
for both single-processor and distributed parallelized sequential systems.
In the latter case, however, the joint work must also be organized, implemented with additional machine instructions and additional execution time, forming an overhead\footnote{The 'weak scaling' approximation neglects this aspect. A many billion USD mistake.}. This extra activity originates from efficiency: one of the processors orchestrates the joint operation, and the others are idly waiting.
At this point, the "\textit{dark performance}" appears: the processing
units are ready to operate and consume power but do not make any payload work.
As discussed in detail in~\textbf{\cite{VeghTemporal:2020}},
the "stealthy nature" of the incremental development of technology made its appearance
unnoticed. However, today, the "idle time" is the primary reason that power consumption is used mainly
for delaying electronic signals~\cite{ClockDistribution:2012} inside our computing systems and
delivering data rather than performing computations~\cite{WhyNotExascale:2014}.

\begin{figure*}[t!]
	\maxsizebox{\textwidth}{!}
	{
		\begin{tabular}{cc}
			\hspace{-.5cm}
			\includegraphics[width=.48\textwidth]
			{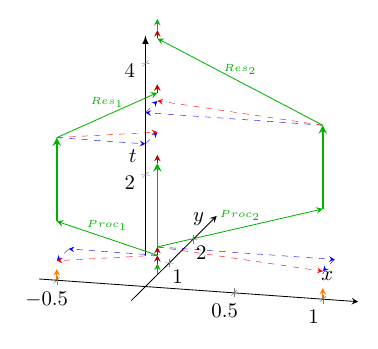}				
			&
			\hspace{-1cm}
			\includegraphics[width=.65\textwidth]
			{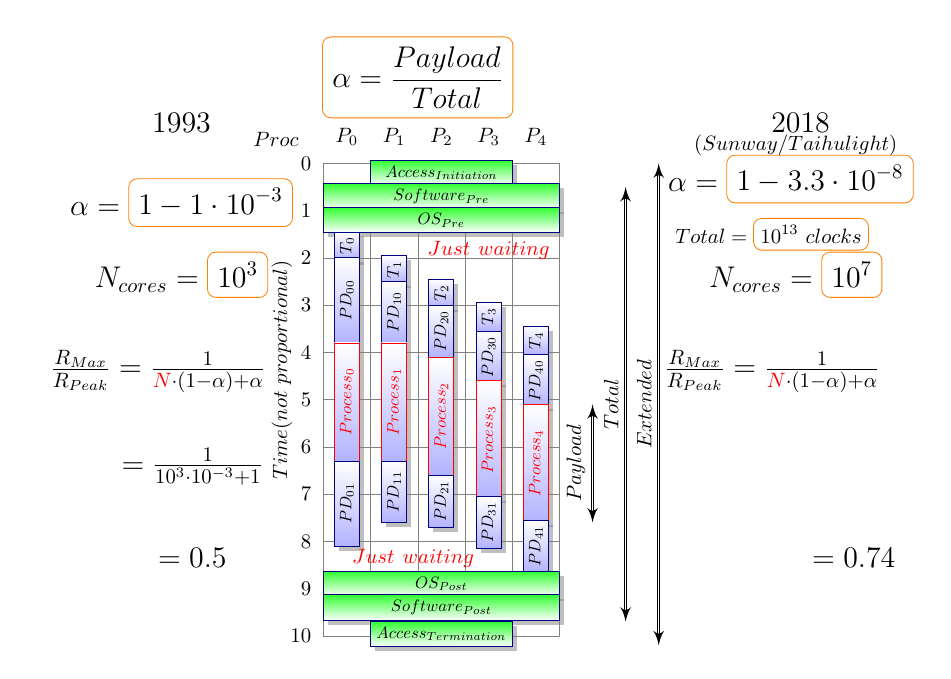}				
		\end{tabular}
	}
	\caption{Left: The time diagram of parallelized sequential operation in time-space~\protect{\textbf{\cite{VeghTemporal:2020}}.
			Right: A conceptual, simplified model~\protect{\cite{VeghScalingANN:2021}} of parallelized 
			sequential computing operations, based on their temporal behavior. 
			Notice the different nature of those contributions, and that
			they have only one common feature: \textit{they all consume time}.
			\label{fig:AmdahlModelAnnotated}}}
\end{figure*}

Amdahl listed~\cite{AmdahlSingleProcessor67} different reasons why losses in "computational load" can occur.
Amdahl's idea enables us to \textit{put everything that cannot be parallelized, i.e., distributed between fellow processing units, into the sequential-only fraction of the task}. For describing the parallelized operation
of sequentially working units, the model depicted in Figure~\ref{fig:AmdahlModelAnnotated} was prepared
(based on the temporal behavior of components, as described in~\textbf{\cite{VeghTemporal:2020}}).
The technical implementations of the different parallelization methods show up a virtually infinite variety~\cite{HwangParallelism:2016}, so here a (by intention) strongly simplified model is presented.
However, the model is general enough to qualitatively discuss  systems working in parallel. We shall neglect the various contributions
as possible in the different cases. Although with some obvious limitations, one can easily convert our model to a technical (quantitative) one by interpreting its contributions in technical terms.
Such technical interpretations also enable us to find out some technical limiting factors of the performance of parallelized computing.

\label{sec:TheModel}

The non-parallelizable (i.e., apparently sequential) part of tasks comprises contributions from \gls{HW}, \gls{OS}, \gls{SW} and \gls{PD}, and also some access time is needed for reaching the parallelized system.
This separation is rather conceptual than strict, although dedicated measurements can reveal their role, at least approximately.
Some features can be implemented in either \gls{SW} or \gls{HW} or shared between them.
Furthermore, some sequential activities
may happen partly parallel with each other.
The relative weights of these contributions are very different for different parallelized systems
and even, within those cases, depend on many specific factors.
In every single parallelization case, a careful analysis is required.
\gls{SW} activity represents what was assumed and discussed by Amdahl as the total sequential fraction\footnote{Although indeed some \gls{OS} activity was included, Amdahl concluded 20~\% \gls{SW} fraction. At that time, the other contributions
	could be neglected apart from \gls{SW} contribution. As shown in Figure~\ref{fig:AmdahlModelAnnotated} and discussed below,  today this contribution became by several orders of magnitude smaller. However, at the same time, the
	number of the cores grew several orders of magnitude larger.}.
Non-determinism of modern \gls{HW} systems~\cite{PerformanceCounter2013}~\textbf{\cite{Molnar:2017:Meas}} also contributes to the non-parallelizable portion of the task: the resulting execution time
of parallelly working processing elements is defined by the slowest unit.

Notice that our model assumes no interaction between processes running on the parallelized system and the necessary minimum: starting and terminating otherwise independent processes, which take parameters at the beginning and return their result at the end. 
It can, however, be trivially extended to more general cases when processes must share some resource (such as a database, which shall provide different records for the different processes), 
either implicitly or explicitly. Concurrent objects have their inherent sequentiality~\cite{InherentSequentiality:2012}. Synchronization and
communication between those objects 
considerably increase~\cite{YavitsMulticoreAmdahl2014} the non-parallelizable portion
(i.e. contribution to $(1-\alpha_{eff}^{SW})$ or $(1-\alpha_{eff}^{OS})$) of the task. 
Because of this effect, in the case of many processors, we must devote special attention to their role in the efficiency of applications on parallelized systems. 

The physical size of the computing system also matters.
A processor connected to the first one with a cable of dozens of meters must wait for several hundred clock cycles.
This waiting is only because of the finite speed of light propagation, topped by their interconnection's latency time and hoppings (not mentioning geographically distributed computer systems, such as some clouds, connected through general-purpose networks).
Detailed calculations are given in~\textbf{\cite{Vegh:2017:AlphaEff}}.

After reaching a certain number of processors, there is no more increase in the payload fraction when adding more processors.
The first fellow processor has already finished its task and is idly waiting, while the last one is still idly waiting for its start command.  We can increase this limiting number by organizing the processors into clusters: the first computer must speak \textit{directly} only to the head of the cluster.
Another way is to distribute the job near to the processing units.
It can happen either inside the processor~\cite{CooperativeComputing2015,DongarraFugakuSystem:2020},
or one can let do the job by the processing units
of a \gls{GPU}\footnote{Notice, however, that any additional actor on the scene increases the latency of computation.}.

This looping contribution is not considerable (and, in this way, not noticeable) at a low number
of processing units (apart from the other contributions). Still, it can dominate at a high
number of processing units. This "high number" was a few dozens
at the time of writing the paper~\cite{ScalingParallel:1993}, today
it is a few millions\footnote{Strongly depends on the workload and the architecture.}.
We consider the effect of the looping contribution as the borderline between first and second-order approximations in modeling the system's payload performance. The housekeeping keeps growing with the number of processors, and in contrast, the system's resulting performance does not increase anymore. The first-order approximation assumes the contribution of housekeeping as constant. The second-order approximation also considers that as the number of processing units grows, housekeeping grows with, gradually becomes the dominating factor of performance limitation, and decreases payload performance.

As Figure~\ref{fig:AmdahlModelAnnotated}
shows, in the parallelized operating mode (in addition to computation, furthermore communication of data between its processing units) \textit{both software
and 
hardware contribute to the execution time}, i.e., they both must be considered in Amdahl's Law. This finding is not new, again: see
\cite{AmdahlSingleProcessor67}.
Figure~\ref{fig:AmdahlModelAnnotated} also shows where is a place to improve computing efficiency.
When combining \gls{PD} properly with
sequential scheduling, one can considerably reduce non-payload time during fine-tuning the system
(see the cases of performance increases of $Sierra$ and $Summit$, a half year after their appearance on the TOP500 list).
Also, mismatching \textit{total time} and \textit{extended measurement time}
(or not making a proper correction) may lead to completely wrong conclusions~\cite{BenchmarkingClouds:2017} as discussed in~\textbf{\cite{Vegh:2017:AlphaEff}}.

\section{Amdahl's Law in terms of our model}\label{sec:AmdahlsLaw}

Usually, Amdahl's law is expressed as 
\vspace{-.3\baselineskip}	
\begin{equation}
S^{-1}=(1-\alpha) +\alpha/N \label{eq:AmdahlBase}
\end{equation}

\noindent where $N$ is the number of parallelized code fragments (or \gls{PU}s), 
$\alpha$ is the ratio of the parallelizable portion to the total,
$S$ is the measurable speedup. From this

\vspace{-.3\baselineskip}	
\begin{equation}
\alpha = \frac{N}{N-1}\frac{S-1}{S} \label{equ:alphaeff}
\end{equation}

\begin{figure}
	\includegraphics[width=\textwidth]{
		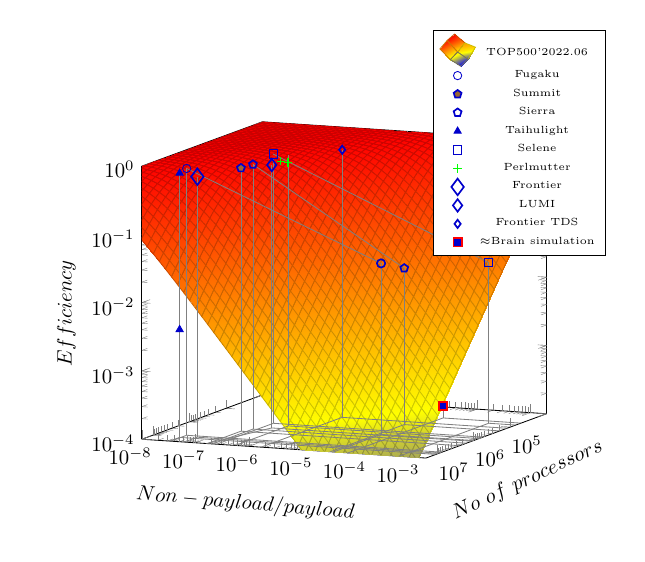}
	\caption{The 2-parameter efficiency surface (in function of parallelization efficiency and number of the processing elements), as concluded from Amdahl's Law (see Eq.~(\ref{eq:soverk})), in first order approximation. Some sample efficiency values for
		some selected supercomputers are shown, measured with benchmarks \gls{HPL} and \gls{HPCG}, respectively. ”\textit{This decay in performance is not a fault of the
			architecture, but is dictated by the limited parallelism}”~\cite{ScalingParallel:1993}
		\label{fig:EffDependence2020Log}
	}
\end{figure}

\noindent
When calculating the speedup, one calculates
\vspace{-.3\baselineskip}	
\begin{equation}
S=\frac{(1-\alpha)+\alpha}{(1-\alpha)+\alpha/N} =\frac{N}{N\cdot(1-\alpha)+\alpha}
\end{equation}
hence the resulting \textit{efficiency} of the system (see Figure~\ref{fig:EffDependence2020Log})
\vspace{-.3\baselineskip}	
\begin{equation}
\boxed{E(\large N,\alpha)} = \frac{S}{N}=\boxed{\frac{1}{\textcolor{red}{\Large N}\cdot(1-\alpha)+\alpha}}= \frac{R_{Max}}{R_{Peak}} \label{eq:soverk}
\end{equation}

This phenomenon itself has been known for decades
\cite{ScalingParallel:1993}, and $\alpha$ is theoretically
established~\cite{Karp:parallelperformance1990}.
Presently, however, the theory was somewhat faded, mainly due to the rapid development of parallelization technology and the increase in single-processor performance.

During the past quarter of a century, the proportion of contributions changed considerably: 
today the number of processors is thousands-fold higher than a quarter of a century ago. 
The growing physical size and the higher processing speed
increased the role of the propagation overhead,
furthermore, the large number of processing units enormously amplified the
role of the looping overhead.
As a side-result of the technological development, \textit{the phenomenon 
	on performance limitation, returned in a technically 
	different form, at a much higher number of processors}.

Through using Eq.~(\ref{eq:soverk}), $E=\frac{S}{N}= \frac{R_{Max}}{R_{Peak}}$ can be equally good for describing the efficiency of parallelization of a setup:
\begin{equation}
\alpha_{E,N} = \frac{E\cdot N -1}{E\cdot (N-1)}\label{eq:alphafromr}
\end{equation}

\noindent 
As we discuss below, except for an extremely high number of processors, we can safely assume that the value of $\alpha$ is independent
from the number of processors in the system. 
Eq.~(\ref{eq:alphafromr}) can be used to derive the value of $\alpha$ from values of parameters $R_{Max}/R_{Peak}$ and the number of cores $N$.

\textit{According to Eq.~(\ref{eq:soverk}), the payload efficiency can be described
	with a 2-dimensional surface}, as shown in Figure~\ref{fig:EffDependence2020Log}. On the surface, we displayed some measured efficiencies of the current top supercomputers to illustrate some general rules. 
Both the \gls{HPL}\footnote{http://www.netlib.org/benchmark/hpl/}
and the \gls{HPCG}\footnote{https://www.epcc.ed.ac.uk/blog/2015/07/30/hpcg}
efficiency values are displayed.
We project the measured values back to the axes to enable the reader to 
compare the corresponding values of their number of processors and
parallelization efficiency.  The \gls{HPL} efficiency sharply decreases with the increasing number of cores in the system.
In the case of unnaturally high (or not provided) \gls{HPCG} efficiency values,
the values are not displayed in the figure.
 The last "world champion" which benchmarked \gls{HPCG} correctly was $Taihulight$.
  
We can divide the measured values into two groups.
The first group comprises measurements where the same number of cores were used in both benchmarks.  For visibility, only the \gls{HPL} projections are displayed,
and on their top, the \gls{HPCG} data point. The general experience showed that the ratio of \gls{HPL}-to-\gls{HPCG} efficiency is about 200-500 \textit{when using the same number of cores}.
 The \gls{HPCG} payload performance reached its "roofline"~\cite{WilliamsRoofline:2009}~\textbf{\cite{VeghScalingANN:2021}} level at a lower number of cores; using all cores would decrease the system's performance because of the higher number of cores.
 This issue is why the members of the second group reduced their number of cores in the \gls{HPCG} benchmark.

The recent trend is that only a tiny fraction (only about 10~\%) of supercomputer's cores  are used in the \gls{HPCG} benchmarking,
while all cores are used in the \gls{HPL} benchmarking.
Beginning with June 2021, the data "Measured cores" are not provided anymore, covering this aspect. The presumable reason is that in this way the measured \gls{HPCG}/\gls{HPL} efficiency ratio gets much higher, providing the illusion that the vast supercomputers became more suitable for real-life tasks. 
The supercomputers based on the newly developed AMD EPYC 7763 64C processors also
 did not provide their \gls{HPCG} efficiency and performance, or they provide an incorrect number of measured cores.

There is an inflection point in the performance:
"\textit{there comes a point when using more \gls{PU}s \dots actually increases the execution time rather than reducing it}"~\cite{ScalingParallel:1993}. We can observe a quick breakdown
of the performance gain as theoretically predicted~\textbf{\cite{VeghScalingANN:2021}} and experimentally measured, see~\cite{AngeloLearningBioinformatics:2014}(Fig.~13) or \cite{NeuralScaling2017}(Fig.~7).
As can be concluded from the figure, increasing the systems' \textit{nominal performance} by order of magnitude, at the same time,
decreases its efficiency (and so: its \textit{payload performance}) by more than an order of magnitude.

\begin{figure*}
		\includegraphics{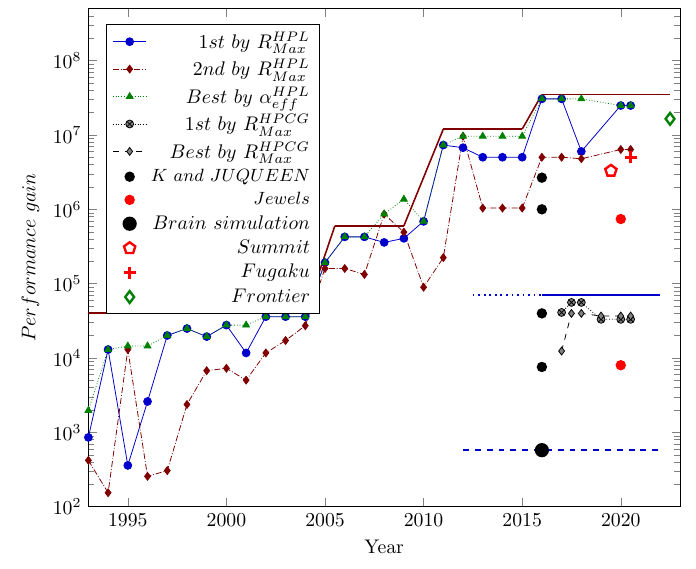}
	\caption{The performance gain of top supercomputers, in function of their year of production. The marks
		display the measured values
		derived using \gls{HPL} and \gls{HPCG} benchmarks, for the TOP3 supercomputers.
		The small black dots mark the performance data of
		supercomputers $JUQUEEN$ and $K$ as of 2014 June,
		for \gls{HPL} and \gls{HPCG} benchmarks, respectively.
		The big black dot denotes the performance value of the system used by \cite{NeuralNetworkPerformance:2018}.
		The red pentagons denote performance gain, measured using half-precision operands.
		The saturation effect can be observed for both
		\gls{HPL} and \gls{HPCG} benchmarks.
		\label{fig:RooflineBrain}
}
\end{figure*}

The goal of the Gordon Bell Prize~\cite{GordonBellPrize:2017} was originally "\textit{to demonstrate a speedup of at least 200 times \textbf{on a real problem}}", and the community noticed a decade ago that the efficacy measured with the benchmark \gls{HPL} and that of the real-life applications started to differ by up to two orders of magnitude.
It would be worth recalling that the new benchmark program \gls{HPCG}~\cite{HPCGB:2015}  was introduced since "\textit{\gls{HPCG} is designed to exercise computational and data access patterns that more closely match a different and broad set of important applications, and to give incentive to computer system designers to invest in capabilities that will have impact on the collective performance of these applications}"~\cite{HPCG_List:2016}.
The present design efforts target building "racing" computers,
and the constructors either do not publish their \gls{HPCG} efficiency or measure it using only a fraction (about 10~\%)
of their cores. The developers aim to produce higher figures by the \gls{HPL} benchmark instead of making higher performance in the real-life-imitating \gls{HPCG} benchmark.  
It is simply misleading to claim that building such racing supercomputers "\textit{will enable scientists to develop critically needed technologies for the country’s energy, economic and national security, helping researchers address problems of national importance that were impossible to solve just five years ago}" and that "\textit{Scientists and engineers from around the world will put these extraordinary computing speeds to work to solve some of the most challenging questions of our era}"~\cite{ORNLFrontier:2022}.

Comments such as "\textit{The \gls{HPCG} performance at only 0.3\% of peak performance
	shows the weakness of the Sunway TaihuLight architecture with slow memory and modest
	interconnect performance}"~\cite{DongarraSunwaySystem:2016}
and "\textit{The \gls{HPCG} performance at only
	2.5\% of peak performance shows the strength of the memory architecture performance}"~\cite{DongarraFugakuSystem:2020}
show that supercomputing experts did not realize, that the efficiency is a(n at least) 2-parameter function, depending on \textit{both} the number of \gls{PU}s in the system and its workload; furthermore, that the workload defines the achievable payload performance.
It looks like the community experienced the effect of the two-dimensional efficiency
but did not want to comprehend its reason, despite the early and clear prediction: "\textit{this decay in performance is not a fault of the
	architecture, but is dictated by the limited parallelism}"~\cite{ScalingParallel:1993}. In excessive systems of modern \gls{HW}, \textit{it is also dictated by the laws of nature}~\textbf{\cite{VeghModernParadigm:2019}}.
Furthermore, we can perfectly describe its dependence by the correctly interpreted Amdahl's Law,
rather than being "empirical efficiency".\index{empirical efficiency}
\index{efficiency!empirical}

Notice two more rooflines. During the development of the interconnection technology, between the years 2004 and 2012, different implementation ideas 
have been around, and they competed for years.
The beginning of the second roofline, around the year 2011,
coincides with the dawn of \gls{GPU}s, interfering with the
effect of the interconnection technology; see also section~\ref{sec:interconnection}.

The top roofline dawned with the appearance of $Taihulight$:
some assistant processors take over part of the duties of the
individual cores, and in this way, one can mitigate the non-payload portion of the
workload. This roofline can be slightly above the possible roofline at the price of using a slightly modified computing paradigm; using cooperating cores.

It is important to notice the two red pentagons in the figure:
they represent the performance gain achieved using half-precision 
operand length. The performance gain is lower than the double precision
equivalent, just because of the increased relative weight of housekeeping,
as discussed in detail in section~\ref{sec:operandlength}.

The projections of efficiency values to axes show that the top few
supercomputers offer similar parallelization efficiency and core number values; both features are required to receive one of the top slots. 
Supercomputers $Taihulight$
and $Fugaku$
are exceptions on both axes. They have the highest
number of cores and the best $HPL$ parallelization efficiency. 
An interesting coincidence is that the processors of both supercomputers have "assistant cores"
(i.e., some cores do not make payload computing; instead, they take over "housekeeping duties").
This solution makes housekeeping in parallel with the payload computing (reduces the sequential portion of the task) and, in this way, decreases the internal latency of processors making payload computing
and increases the system's efficiency. They both use a "light-weight operating system" (and so do $Fugaku$, $Summit$ and $Sierra$) to reduce processor latency. This efficiency, of course, requires executing several
floating instructions per clock cycle. That mode of operation gets more and more challenging
for the interconnection, delivering data to and from data processing units.
Notice also in their cases the role of "near" memories: as explained in~\textbf{\cite{VeghTemporal:2020}},
the data delivery time considerably increases the "idle time" of computing.
This idle time is why $Fugaku$, with its cleverly placed $L_2$ cache memories,
can be more effective when measured with \gls{HPL}. However, this trick
does not work in the case of $HPCG$ because its "sparse" computations
use those cache memories ineffectively. We expect the "true" $HPCG$ efficiency of
$Fugaku$ to be between the corresponding values of 
$Summit$ and $Taihulight$.

The newly (as of 2022) developed AMD EPYC 7763 64C processors did not cause
a revolution. On the one side, they introduced a performance jump, quite similarly to the appearance of the NVidia accelerators some years ago. On the other side, their internal latency 
allowed to achieve a somewhat higher performance gain.
As the data point representing $Frontier$ shows, 
their performance gain is slightly higher than that of systems
without acceleration and considerably higher than that of
systems having accelerators with non-shared data space.
However, they did not require developing new theoretical approach.

The processors of $Taihulight$ comprise cooperating cores~\cite{CooperativeComputing2015}.
The direct core-to core transfer uses a (slightly) different computing paradigm: 
the processor cores explicitly assume the presence of another core, and in this way, their
effective parallelism becomes much better, see also  Fig.~\ref{fig:InterconnectionVsPerformance}.
On that figure, this data 
and the ones using shorter operands ($Summit$ and $Fugaku$) result in performance values above the limiting line.
However, reducing the loop count by internal clustering (in addition to the "hidden clustering",
enabled by its assistant cores)
and exchanging data without using the global memory  works only for the 
$HPL$ case, where the contribution of \gls{SW} is low.
The poor value of $(1-\alpha_{eff}^{HPCG})$ is not necessarily a sign of
architectural weakness~\cite{NSA_DOE_HPC_Report_2016}:  $Taihulight$ comprises about four times more cores
than $Summit$ and performs the \gls{HPCG} benchmark with ten-fold more cores. 
Given that \gls{HPCG} mimics "real-life" applications, one can conclude that for practical purposes, only systems comprising a few hundred thousand cores\footnote{Assuming good interconnection. This limiting number is much less for computing systems
	connected with general-purpose networks, such as some high-performance clouds. Similar is the case with 
real-life applications needing more communication.}
shall be built, see also section~\ref{sec:accelerator}. More cores contribute only to the "dark performance".

According to Eq.~(\ref{eq:soverk}), efficiency can be interpreted
in terms of $\alpha$ and $N$,
and the efficiency of a parallelized sequential computing system
can be calculated as

\begin{equation}
P(N,\alpha) = \frac{N\cdot P_{single}}{{N\cdot \left(1-\alpha\right)+\alpha}}\label{eq:Ppayload}
\end{equation}

\noindent This simple formula explains why \textit{the payload performance 
	is not a linear function of the nominal performance} and why
in the case of very good parallelization ($(1-\alpha)\ll1$)
and low $N$, this nonlinearity cannot be noticed.

The value of $\alpha$, however, can hardly be calculated for the present
complex HW/SW systems from their technical data (for a detailed discussion see~\textbf{\cite{VeghScalingANN:2021}}). We can follow two ways to estimate their value of $\alpha$. One way is to calculate $\alpha$ for existing supercomputing systems
(making "computational experiments"\cite{AmdalVsGustafson96}) applying Eq.~(\ref{eq:alphafromr}) to data in TOP500 list~\textbf{\cite{VeghModernParadigm:2019}}. This way provides a lower bound, already achieved,
for $(1-\alpha)$. Another way round is to consider contributions of different origins, see section~\ref{sec:Contributions}, and to calculate the high limit of the value of $(1-\alpha)$,
that the given contributions alone do not enable to exceed (provided
that that contribution is the dominant one). It gives us good confidence in the reliability of the parameters that the values derived in these two ways differ only by up to a factor of two. At the same time, this also means that technology is already
very close to its theoretical limitations.

Notice that the "algorithmic effects" -- such as dealing with "sparse" data structures
(which affects cache behavior, which will have a growing importance
in the age of "real-time everything", "big data", and "neural networks") or
communication between parallelly running threads, such as returning results repeatedly to the main thread
in an iteration (which greatly increases the non-parallelizable fraction in the main thread) -- 
manifest through the \gls{HW}/\gls{SW} architecture,
and we can hardly separate them.
Also notice that there are one-time and fixed-size contributions, such as utilizing time measurement facilities or calling system services. Since $\alpha_{eff}$ is a \textit{relative} merit, the \textit{absolute}
measurement time shall be large. When utilizing efficiency data from measurements
dedicated to some other goal, we must exercise proper caution with the interpretation and accuracy
of those data.

The 'right efficiency metric'~\cite{EfficiencyMetric:2012} has always been a question (for a summary see cited references in~\cite{HSWscalable2012})
when defining efficient supercomputing.  The present discussion aims to discover the inherent limitations of parallelized sequential computing
and provide numerical values.
For this goal, the 'classic interpretation'~\cite{AmdahlSingleProcessor67,ScalingParallel:1993,Karp:parallelperformance1990} of performance
was used, in its original spirit. We scrutinized the contributions mentioned in those papers
and revised their importance under current technical conditions.

\section{The effect of different contributions to $\alpha$}\label{sec:Contributions}
Theory can display data from systems with any
contributors with any parameters. From measured data, we can conclude only the sum of all contributions,
although dedicated measurements can reveal the
value of separated contributions experimentally.
The publicly available
data enable us to conclude mainly of limited validity.

\subsection{Estimating different limiting factors of $\alpha$}\label{sec:estimatingfactors}

The estimations below assume that the actual contribution is the dominating one, which defines the achievable performance alone. This situation is usually not the case in practice (the convolution of different contributions is limiting), 
but this approach enables us to find out the limiting $(1-\alpha)$ values for all contributions.

In the systems implemented in \gls{SPA}~\cite{AmdahlSingleProcessor67} as parallelized sequential systems,
the life begins and ends in one such sequential subsystem, see also Fig.~\ref{fig:AmdahlModelAnnotated}.
In large parallelized applications, running on general-purpose supercomputers,
initially and finally, only one thread exists,
i.e., the minimal necessary non-parallelizable activity is to fork the
other threads and join them again.

With the present technology, no such actions can be shorter than one processor clock period\footnote{
This statement is valid even if some parallelly working units can execute more than one instruction in a clock period.
One can take these two clock periods as an ideal (but not realistic) case. However, the actual limitation will inevitably be (much) worse than the one calculated for this idealistic case. The exact number of clock periods depends on many factors, as discussed below.}.
The theoretical absolute minimum value of the non-parallelizable portion of the task will be given as the ratio of the time of the two clock periods to the total execution time. 
The latter time is a free parameter in describing efficiency. That is, the value of effective parallelization $\alpha_{eff}$ \textit{depends
on the total benchmarking time} (and so does the achievable parallelization gain, too).

This dependence, of course, is well known for supercomputer scientists. For measuring the efficiency with better accuracy (and also for producing better $\alpha_{eff}$ values),
one uses hours of execution times in practice. In the case of benchmarking supercomputer
$Taihulight$~\cite{DongarraSunwaySystem:2016},  13,298 seconds \gls{HPL} benchmark runtime was used; on the 1.45 GHz processors it means 
$2*10^{13}$ clock periods. The inherent limit of  $(1-\alpha_{eff})$ at such benchmarking time
is $10^{-13}$ (or, equivalently, the achievable performance gain is $10^{13}$).
For simplicity, in the paper, 1.00 GHz processors (i.e., 1~ns clock cycle time) will be assumed.

Supercomputers are also distributed systems.
In a stadium-sized supercomputer, we can assume a distance between processors (cable length) up to about 100 m. The net signal round trip time is ca. 
$10^{-6}$ seconds, or $10^{3}$ clock periods, i.e., in the case of a finite-sized supercomputer, the performance gain cannot be above $10^{10}$,
only because of the physical size of the supercomputer.
The presently available network interfaces have 100\dots200 ns latency times and sending a message between processors 
takes time in the same order of magnitude, typically 500 ns.
This timing also means that \textit{making better interconnection 
is not a bottleneck in enhancing performance}.
This statement is also underpinned by the discussion in section~\ref{sec:interconnection}. However, sharing data instead of copying them, can result in demonstrative performance improvements, see the case of $Frontier$.

These predictions enable us to assume that the presently achieved value of $(1-\alpha_{eff})$ could also persist for roughly a hundred times more cores. 
However, another major issue arises from the computing principle 
\gls{SPA}: the first processor can address only one processor at a time. As a consequence, at least as many 
clock cycles are to be used for organizing the parallelized work as many addressing steps are required.
This number equals the number of cores 
in the supercomputer, i.e., the addressing operation in supercomputers in the TOP10 positions typically needs clock cycles in the order of
$5*10^{5}$\dots$10^{7}$;
degrading the value of $(1-\alpha_{eff})$ into the range
$10^{-6}$\dots$2*10^{-5}$.
One can use two tricks to mitigate the number of the addressing steps: 
either cores are organized into \textit{cluster}s
as many supercomputer builders do, or at the other end, the processor itself can
take over the responsibility of addressing its cores~\cite{CooperativeComputing2015}.
In function of the actual construction, the reducing factor
of clustering of those types can be in the range $10^{1}$\dots$5*10^{3}$,
i.e the resulting value of $(1-\alpha_{eff})$ is expected to be around $10^{-7}$.

As Eq.~(\ref{eq:Ppayload}) suggests,
the trivial way to increase systems' performance is to improve a single processor's performance; for example, combining the \gls{PU}s with accelerators. Various implementations of the idea exist, but 
"\textit{What is initially perplexing is that accelerators do not dominate the architectures in the Top500, given the performance and price/performance advantages they offer as compute engines}"~\cite{NextPlatformHPCExascaleBarrier:2022}.
This situation seems to change: now the Frontier has taken over the leading position, and presumably, other supercomputers will be built with its technology. The primary reason for the change is that sharing data (mainly at $L_1$ and $L_2$ levels), as opposed to copying data between the separated address fields of \gls{PU} and \gls{GPU}, dramatically reduces the internal latency of the \gls{PU}s, and in the case of "racing" supercomputers, this non-payload contribution can be decisive, see section~\ref{sec:WorkloadOnPerformance}. A critical difference between the previous and recent designs of \gls{GPU}s is that 
"\textit{The two GK110 GPUs in the K80s were not interconnected and could not share data; the two Aldebaran GCDs are and can.}"
\cite{NextPlatformAMDAldebaran:2022}".  
The consequence of the lower latency is a higher performance gain and, due to it, a higher absolute performance. The former \gls{GPU}s, because of their higher latency due to copying between address spaces, could not get into the top group of supercomputers.

\begin{figure}[h]
	\includegraphics[width=1.05\columnwidth]
	{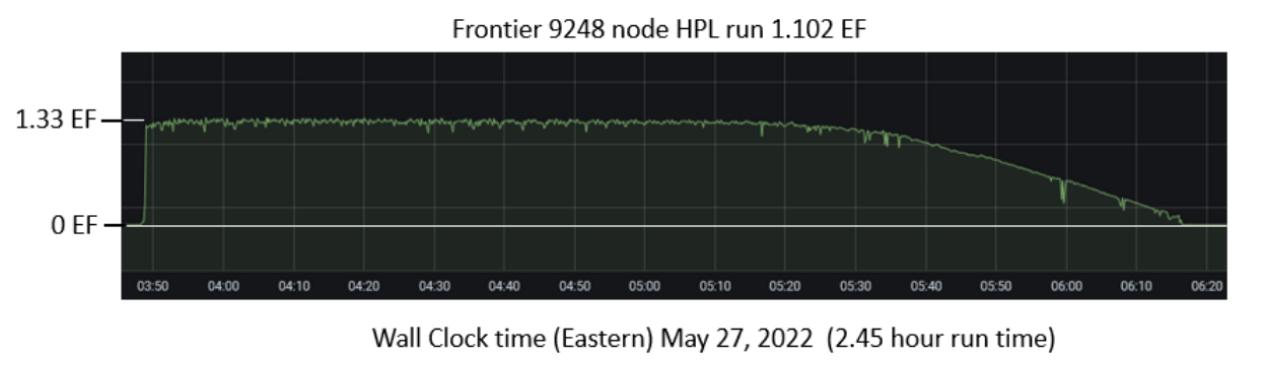}\vspace{-\baselineskip}
	\caption{The time dependence of the number of floating point operations performed when running benchmark \gls{HPL} on supercomputer Frontier\label{fig:FrontierHPLDependence}. Reproduced from~\cite{FrontierHPCwire:2022}}
\end{figure}

\cite{NextPlatformFrontierStepByStep:2022} introduces an unexplained claim: "\textit{As configured for the Linpack run, the Frontier system had a power draw of 21.1 megawatts, but \dots
	when the Frontier machine is first set loose on the Linpack, it draws an extra 15 megawatts of power as it is starts the job.}" Together with Fig.~\ref{fig:FrontierHPLDependence} one can understand why. 

At the very beginning and end of computation, the operands shall be distributed and results collected, respectively. 
This non-payload contribution grows with the number of cores.
As discussed above, several ideas are used to reduce its time: from the very common "clustering" to using assistant cores~\cite{CooperativeComputing2015}.
Frontier introduced a revolutionary solution for that task. 
As the left side of Fig.~\ref{fig:FrontierHPLDependence} illustrates, for about 3 minutes, Frontier, at least its monitored part, does not perform floating operations. After that period, all operands suddenly appear in their place, and the supercomputer can make its computations at full speed; apparently, no time is spent transferring the operands to the peer cores. This task is what Frontier uses the extra 15~MW power for; for this short period, a hidden second specialized supercomputer starts up, with the only mission to deliver the operands to their place. From the viewpoint of the coordinator core, it is read-only access that can be massively parallelized. However, it needs a terrible parallel bus capacity and electric power to drive millions of \gls{I/O} ports\footnote{Normally, communication needs assistance from the \gls{OS}.}.

The right side of Fig.~\ref{fig:FrontierHPLDependence} illustrates that when the benchmark computation \gls{HPL} finishes, it takes about 40 minutes to collect the data from its nearly 9M cores. It enables us to estimate also that the value of $T_T$ (the time needed for a single transfer) is about $200,000\ ns$~\cite{FrontierHPCwire:2022}; a value unexpectedly high compared to Slingshot's advertised 200 Gb/sec transfer bandwidth.  The issue shows a very poor bus utilization. The datasheet value does not consider the arbitration time and data transmission time. This finding is not unprecedented:
\cite{NextPlatformHPCExascaleBarrier:2022} mentions that
"\textit{IBM and Nvidia had issues with the NVLink coherent fabric between the CPUs and GPUs in Summit, and that machine did not get 200 Gb/sec InfiniBand as was hoped when it was installed.}" Yes, the "proxy" core and the separated CPU/GPU address space increased the node's internal latency (its non-payload contribution).

The figure also shows that the submitted computing performance is averaging the momentary values for the period shown in the figure.  Without the "phantom supercomputer" (i.e., having \textit{two} ramps), the performance would be about 0.8~EFlops (that is, \textit{less} than the magic 1 EFlops), and with a sufficiently long measurement time, it could approach 1.3~EFlops.
Notice that the "ramp" on the right side would occur 
several times in \gls{HPCG}-type computations (and, of course, in all real-life computations) would keep the average computing performance low. To prevent this, one can reduce the number of the measured cores to, say, 10\%, in which case the ramp's length is about 4 minutes, and in that way, much better efficiency and speedup can be reported.

The 'end of computation' situation means that the orchestrating core must collect the result from all peer cores. It imitates the case when in an  \gls{AI} workload a 'neuron' must collect the results from all of its (interested) peers, and all neurons want to make that action at the same time, using the same single bus.
For a \gls{AI}-type application, the increased level of communication  keeps \gls{PU}s waiting for the single high-speed bus so that the computed results "\textit{are processed as they come in}",
and to reduce the apparent overload,
"\textit{are dropped if the
	receiving process is busy over several delivery cycles}"~\cite{NeuralNetworkPerformance:2018}.
Actually, using a single high-speed bus excludes the chance
of achieving a real-time simulation speed.
This method of implementation is why
"\textit{artificial intelligence, \dots it's the most disruptive workload from an I/O pattern perspective}"\footnote{ https://www.nextplatform.com/2019/10/30/cray-revamps-clusterstor-for-the-exascale-era/}.

One must also use an operating system for protection and convenience.
If fork/join is executed by the \gls{OS} as usual, because of the needed context switchings $2*10^{4}$~\cite{Tsafrir:2007,armContextSwitching:2007} clock cycles are used, rather than 2 clock cycles considered in the idealistic case. \textit{The derived values are correspondingly by four orders of magnitude different}, that is, the absolute limit is $\approx~5*10^{-8}$, on a zero-sized supercomputer. This value is somewhat better than the limiting value derived above, but it is close to that value and surely represents a considerable contribution.
This limitation is why a growing number of supercomputers uses "light-weight kernel" or runs its actual computations in kernel mode; a method of computing that we can use only with well-known benchmarks. Frontier's 'phantom supercomputer' eliminates also
this \gls{OS} contribution.

However, this optimistic limit assumes that the processor can access its instructions in one clock cycle. It is usually not the case, but it seems to be a good approximation. 
On the one side, 
even a cached instruction in the memory needs about five times more
access time and the time required to access 'far' memory is roughly 100 times longer.
Correspondingly, the optimistic achievable performance gain values shall be scaled down by a factor of $5\dots100$. 
A considerable part of the difference between efficiencies $\alpha_{eff}^{HPL}$
and $\alpha_{eff}^{HPCG}$ can be attributed to a different 
cache behavior because of the 'sparse' matrix operations.

\subsection{The effect of workload\label{sec:WorkloadOnPerformance}}

\begin{figure}
\includegraphics[width=\textwidth]{
	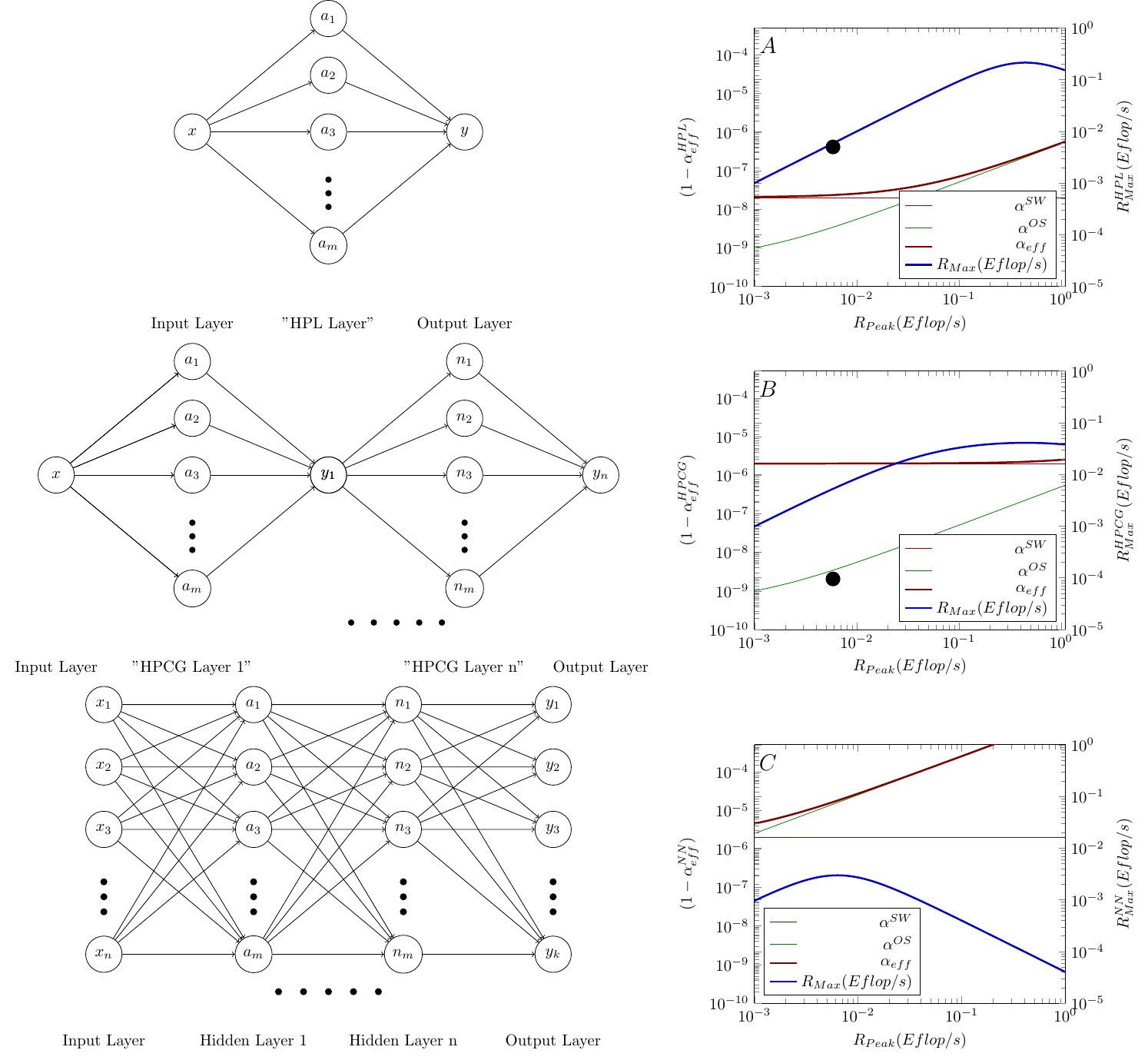}
\caption{The figure explains how the different communication/computation intensities of applications
	lead to different payload performance values in the same supercomputer system. Left column: models of the computing intensities for different benchmarks. Right column: the corresponding payload performances and $\alpha$ contributions in function of the nominal performance of a fictive supercomputer ($P=1Gflop/s$ @ $1GHz$).
	The blue diagram line refers to the right hand scale ($R_{Max}$ values), all others ($(1-\alpha_{eff}^{X})$ contributions) to the left hand scale. The figure is purely illustrating the concepts; the displayed numbers are somewhat similar to real ones. The performance breakdown shown in the figures were experimentally measured by~\protect{\cite{ScalingParallel:1993}}, ~\protect{\cite{NeuralScaling2017}}(Figure~7)
	and~\protect{\cite{NeuralNetworkPerformance:2018}}(Figure~8).		
	\label{fig:AlphaContribBenchmark}		
}
\end{figure}

The overly complex Figure~\ref{fig:AlphaContribBenchmark}
attempts to explain the phenomenon, why and how the performance of a
supercomputer configuration depends on the application it runs.
The non-parallelizable fraction (denoted in the figure by $\alpha_{eff}^{X}$) of the computing task comprises components $X$ of different origins. As we already discussed and was noticed decades ago, "\textit{the inherent communication-to-computation ratio in a
parallel application is one of the important determinants
of its performance on any architecture}"~\cite{ScalingParallel:1993}, 
suggesting that \textit{communication can be a dominant contribution in system’s performance}.
Figure~\ref{fig:AlphaContribBenchmark}.A displays a case with minimum communication, and Figure~\ref{fig:AlphaContribBenchmark}.B a case with moderately increased communication
(corresponding to real-life supercomputer tasks).
As the nominal performance increases linearly and the payload performance decreases inversely with the number of cores, 
at some critical value, where an inflection point occurs, the
resulting payload performance starts to drop.
The resulting non-parallelizable fraction sharply decreases the efficacy
(or, in other words: performance gain or speedup) of the system~\textbf{\cite{VeghPerformanceWall:2019,VeghBrainAmdahl:2019}}.
The effect was noticed early~\cite{ScalingParallel:1993}, under different
technical conditions, but somewhat faded due to successes of development of parallelization technology.

Figure~\ref{fig:AlphaContribBenchmark}.A illustrates the behavior measured with \gls{HPL} benchmark. The looping contribution becomes remarkable around 0.1~Eflops, and breaks down payload performance (see Figure~1 in~\cite{ScalingParallel:1993}) when approaching 1~Eflops.
In Figure~\ref{fig:AlphaContribBenchmark}.B, the behavior measured with benchmark \gls{HPCG} is displayed. The application's contribution (brown line) is much higher than in the previous case. The looping contribution (thin green line) is the same as above. 
Consequently, the achievable payload performance is lower, and the payload performance breakdown is softer in the case of real-life tasks.

Figure~\ref{fig:RooflineBrain} depicts the experimental equivalent of Figure~\ref{fig:AlphaContribBenchmark}.
Given that no dedicated measurements exist, it is hard to compare theoretical prediction directly to measured data. However,
the impressive and quick development of interconnecting technologies provides a helping hand.

In the \gls{HPL} class, the communication intensity is the lowest possible one:
the computing units receive their task (and parameters) at the beginning
of the computation, and they return their result at the very end.
The core orchestrating their work must only deal with the fellow cores in these periods, so their communication intensity is proportional to the number of cores in the system. Notice the need to queue requests at the task's beginning and end.

In the \gls{HPCG} class, iteration takes place: the cores return the result of one iteration to the coordinator core, which makes sequential operations: not only receives and re-sends parameters but also needs to compute new parameters before sending them to the fellow cores. The program repeats the process several times. As a consequence, the \textit{non-parallelizable fraction of the benchmarking time grows proportionally to the number of iterations and the size of the problem}.
The effect of that extra communication decreases the achievable performance roofline~\cite{WilliamsRoofline:2009}: as shown in Fig.~\ref{fig:RooflineBrain}, the \gls{HPCG} roofline is about 200 times smaller than the \gls{HPL} one, as discussed in section~\ref{sec:AmdahlsLaw}.
 Turning the memory into (partly) active elements, using different 'coherence' solutions such as OpenCAPI~\cite{OpenCAPI:2019} or data sharing at levels $L_1$ and $L_2$, can mitigate this effect. See also section~\ref{sec:accuracy}.

\subsection{The effect of interconnection}\label{sec:interconnection}

As discussed above, in a somewhat simplified view, 
we can calculate the resulting performance using the contributions to $\alpha$ as

\begin{equation}
P(N,\alpha) = \frac{N\cdot P_{single}}{\textcolor{red}{N}\cdot \left(1-\textcolor{red}{\alpha_{Net} -\alpha_{Compute}} -\alpha_{Others}
\right)+\approx 1}
\end{equation}

\noindent 
That is, we must handle two of the contributions with emphasis.
The theory provides values for contributions of interconnection and calculation separately. 
Fortunately, the public database TOP500~\cite{TOP500}
also provides data measured under conditions considerably similar
to measure the 'net' interconnection contribution.
Of course, the measured data reflect the sum of contributions of all components.
However, as will be shown below, in the total contribution, those mentioned contributions dominate,
and all but the contribution from networking are (nearly) unchanged,
so the \textit{difference} of measured $\alpha$ can be directly compared to the \textit{difference} of the corresponding sum of calculated
$\alpha$ values, although here only qualitative agreement can be expected.

\begin{figure}
\hspace{-.7cm}\includegraphics[width=1.1\textwidth]{
	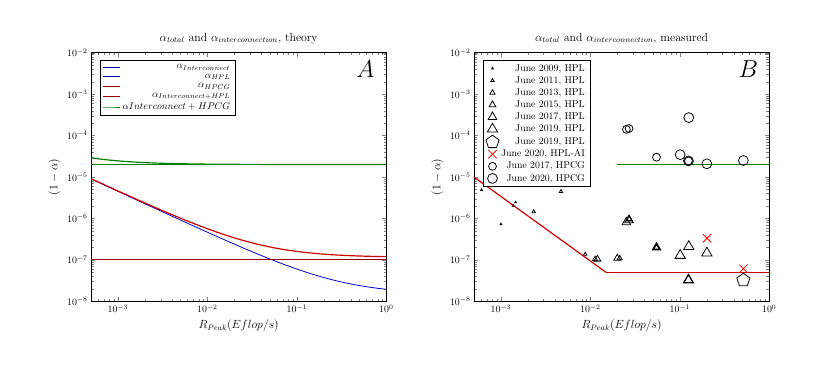}	
\caption{The effect of changing the dominating contribution.
	The left subfigure shows the theoretical estimation,
	the right subfigure the corresponding measured data, as derived from
	public database TOP500~\cite{TOP500} (only
	values for the first four supercomputers are shown).
	When the contribution from interconnection
	drops below that of computation, the value of 
	$(1-\alpha)$ (and the performance gain) get saturated.
	The red 'x' marks denote half-precision values.
	\label{fig:InterconnectionVsPerformance}
}
\end{figure}

Both the quality of the interconnection and the nominal performance are a parametric function of their construction time. One can assume on the
theory side that (in a limited period), interconnection contribution
was changing in the function of nominal performance as shown in Figure~\ref{fig:InterconnectionVsPerformance}A.
The other primary contribution is assumed to be computation\footnote{This time also accessing data ("accessing data" is included). The idly waiting means increased non-payload contribution.} itself.
Benchmark computation contributions for  \gls{HPL} and \gls{HPCG} are very different, so the sum of the respective component plus the interconnection component is also very different.
Given that at the beginning of the considered period,
the contribution from the \gls{HPCG} calculation and that of the
interconnection was in the same order of magnitude, their sum
only changed marginally (see the upper diagram lines), i.e., the measured performance improved only slightly.
Because the benchmark \gls{HPCG}  is communication-bound (and so are real-life programs), 
their efficiency would be an order of magnitude worse.
The reason is Eq.~(\ref{eq:soverk}): when supercomputers use all of their cores,
their achievable performance  is not higher (or maybe even lower),
only their power consumption is higher (and the calculated efficiency is lower).
As predicted: "\textit{scaling thus put larger machines at an inherent
disadvantage}"~\cite{ScalingParallel:1993}.
The cloud-like supercomputers have a disadvantage
in the \gls{HPCG} competition: the Ethernet-like operation
results in relatively high $(1-\alpha)$ values.

The case with \gls{HPL} calculation is drastically different (see the lower diagram lines).
In this case, at the beginning of the considered period, the contribution from interconnection is
very much larger than that from the computation. Consequently, the sum of these two contributions changes sensitively as the speed of the interconnection
improves. As soon as the contribution from interconnection
decreases to a value comparable with that from the computation,
the decrease of the sum slows down considerably,
and \textit{further improvement of interconnection causes
only marginal decrease} in the value of the resulting $\alpha$
(and so only a marginal increase in the payload performance).

The measured data enable us to draw the same conclusion,
but one must consider that here multiple parameters may have
changed. Their tendency, however, is surprisingly straightforward.
Figure~\ref{fig:InterconnectionVsPerformance}.B is actually a 2.5D diagram: the size of marks is proportional
to the time passed since the beginning of the considered period.
A decade ago, the speed of interconnection gave a significant contribution
to $\alpha_{total}$. Enhancing it drastically in the past few years increased systems' efficacy. At the same time, because of the stalled single-processor performance, the other technology components changed only marginally.
The computational  contribution to $\alpha$ from benchmark \gls{HPL}
remained constant in the function of time, so the quick improvement of the interconnection technology resulted 
in a quick decrease of $\alpha_{total}$, and the relative weights of
$\alpha_{Net}$ and $\alpha_{Compute}$ reversed.
\textit{The decrease in the value of $(1-\alpha)$ can be considered as the
result of the decreased contribution from interconnection.}

However, the total $\alpha$ contribution decreased considerably \textit{only} until $\alpha_{Net}$ reached
the order of magnitude of $\alpha_{Compute}$. 
This match occurred in the first 4-5 years of the period shown in Figure~\ref{fig:InterconnectionVsPerformance}B:
the sloping line is due to the enhancement of interconnection.
Then, the two contributors changed their role, and the constant contribution due to computation started to dominate,
i.e., the total $\alpha$ contribution decreased only marginally.
As soon as the computing contribution took over the dominating role, the value of $\alpha_{total}$
did not fall anymore: all measured data remained above that value of $\alpha$.
Correspondingly, the payload performance improved only marginally (and due to factors other than interconnection).

At this point, as a consequence of that the dominating contributor changed, it was noticed that the efficacy of benchmark \gls{HPL} and the efficacy of real-life applications started to differ by up to two orders of magnitude.
Because of this, a new benchmark program \gls{HPCG}~\cite{HPCGB:2015} was introduced, since "\textit{\gls{HPCG} is designed to exercise computational and data access patterns that more closely match a different and broad set of important applications
}"~\cite{HPCG_List:2016}.

Since the major contributor is the computing itself, the different benchmarks contribute differently, and since that time the
"\textit{supercomputers have two different efficiencies}"~\cite{DifferentBenchmarks:2017}.
Yes, if the dominating $\alpha$ contribution (from the benchmark calculation) is different, then the same computer shows different efficiencies
in the function of the computation it runs.
Since that time, the interconnection provides less contribution than the computation of the benchmark. Due to that change, enhancing the interconnection contributes mainly to the \textit{dark performance}, rather than to the \textit{payload performance}.
For the goals of exascale computing, the efforts (and expenses) to enhance interconnection seem to be rather pointless. For today, even in the benchmark \gls{HPL}, the sequential contribution of interconnection is only a tiny fraction of the computation itself; for solving real-life problems, enhanced interconnection makes no noticeable difference. The fundamental issue is that the need for communication (including "sparse" computations) blocks computing performance.

\begin{figure*}
	\maxsizebox{\textwidth}{!}
	{
		\begin{tabular}{rr}
			\tikzset{mark options={mark size=2, line width=.5pt}}
			\begin{tikzpicture}
			\begin{axis}[%
			legend style={
				cells={anchor=west},
				legend pos={north west},
			},
			cycle list name={my color list},
			xmin=0, xmax=51,
			ymin=0, ymax=200, 
			ylabel={Processor performance (Gflop/s)},
			xlabel={Ranking  by $HPL$} ,
			scatter/classes={%
				A={ mark=diamond*,  draw=webgreen},
				N={ mark=triangle*,  draw=webblue},
				G={ mark=square*,  draw=webred}
			}
			]
			\addplot[scatter,only marks,%
			scatter src=explicit symbolic]%
			table[meta=label] {
				x y label
				2	17.6	A
				42  17.4    A
				50  17.6    A
			};
			\addlegendentry{Accelerated}

			\addplot[scatter,only marks,%
			scatter src=explicit symbolic]%
			table[meta=label] {
				x y label
				1	11.78	N
				5  12.8   N
				6  44.8    N
				7  44.8    N
				8  16.1  N
				9  12.8  N
				10  36.8  N
				11  33.6  N
				12  52.9  N
				13  66.96 N
				14  44.8  N
				15  29.5  N
				16  41.6  N
				17  40.0  N
				18  36.8  N
				19  30.4  N
				20  18.4  N
				21  12.8  N
				22  36.8  N
				23  12.8  N
				24  36.8  N
				25  33.6  N
				26  33.6  N
				29  26.8  N
				31  31.6  N
				32  21.6  N
				34  35.2  N
				35  21.6  N
				36  41.6  N
				37  33.6  N
				38  33.6  N
				41  35.4  N
				43  36.8  N
				44  41.6  N
				45  41.6  N 
				46  31.6  N
				47  40    N
				48  35.2  N    	
			};
			\addlegendentry{Non-accelerated}
			
			\addplot[scatter,only marks,%
			scatter src=explicit symbolic]%
			table[meta=label] {
				x y label
				3	70.0	G
				5   48.4    G
				27   84.2    G
				28   84.2    G
				30   83.9    G
				33   36.7    G
				39   79.4    G
				40   25.2    G
				49   69.4    G
			};
			\addlegendentry{GPU-accelerated}
			
			\addplot+[ mark=diamond,  draw=webgreen] table[y={create col/linear regression={y=Y}},
			meta=label,    /pgf/number format/read comma as period
			] {
				x Y label
				2	17.6	A
				20  18.4    A
				50  17.6    A
			};
			\addlegendentry{Regression of accelerated}
			
			\addplot+[ mark=triangle,  draw=webblue] table[y={create col/linear regression={y=Y}},
			meta=label,    /pgf/number format/read comma as period
			] {
				x Y label
				1	11.78	N
				5  12.8   N
				6  44.8    N
				7  44.8    N
				8  16.1  N
				9  12.8  N
				10  36.8  N
				11  33.6  N
				12  52.9  N
				13  66.96 N
				14  44.8  N
				15  29.5  N
				16  41.6  N
				17  40.0  N
				18  36.8  N
				19  30.4  N
				20  18.4  N
				21  12.8  N
				22  36.8  N
				23  12.8  N
				24  36.8  N
				25  33.6  N
				26  33.6  N
				29  26.8  N
				31  31.6  N
				32  21.6  N
				34  35.2  N
				35  21.6  N
				36  41.6  N
				37  33.6  N
				38  33.6  N
				41  35.4  N
				43  36.8  N
				44  41.6  N
				45  41.6  N 
				46  31.6  N
				47  40    N
				48  35.2  N    	
			};
			\addlegendentry{Regression of nonaccelerated}
			
			\addplot+[ mark=square,  draw=webgreen] table[y={create col/linear regression={y=Y}},
			meta=label,    /pgf/number format/read comma as period
			] {
				x Y label
				3	70.0	G
				5   48.4    G
				27   84.2    G
				28   84.2    G
				30   83.9    G
				33   36.7    G
				39   79.4    G
				40   25.2    G
				49   69.4    G
			};
			\addlegendentry{Regression of GPU accelerated}
			
			\end{axis}
			\end{tikzpicture}
			
			&
			\tikzset{mark options={mark size=2, line width=.5pt}}
			\begin{tikzpicture}
			\begin{axis}[%
			legend style={
				cells={anchor=west},
				legend pos={south east},
			},
			cycle list name={my color list},
			xmin=0, xmax=51,
			ymin=1e-9, ymax=5e-5, 
			xlabel={Ranking  by $HPL$} ,
			ylabel={$(1-\alpha_{eff}^{HPL})$},
			ymode=log,
			scatter/classes={%
				A={ mark=diamond*,  draw=webgreen},
				N={ mark=triangle*,  draw=webblue},
				G={ mark=square*,  draw=webred}
			}
			]
			\addplot[scatter,only marks,%
			scatter src=explicit symbolic]%
			table[meta=label] {
				x y label
				2	1.99e-7	A
				42  1.72e-6    A
				50  2.77e-6    A
			};
			\addlegendentry{Accelerated}
			
			\addplot[scatter,only marks,%
			scatter src=explicit symbolic]%
			table[meta=label] {
				x y label
				1	3.27e-8	N
				5  1.10e-7   N
				6  1.59e-6    N
				7  1.51e-6    N
				8  1.04e-7  N
				9  2.19e-7  N
				10  1.22e-6  N
				11  6.4e-7  N
				12  3.64e-6  N
				13  4.03e-6 N
				14  3.06e-6  N
				15  8.05e-7  N
				16  2.75e-6  N
				17  1.68e-6  N
				18  1.56e-6  N
				19  1.22e-6  N
				20  1.64e-6  N
				21  3.76e-7  N
				22  7.84e-7  N
				23  4.38e-7  N
				24  2.25e-6  N
				25  6.11e-7  N
				26  6.11e-7  N
				29  3.04e-6  N
				31  9.32e-7  N
				32  2.45e-6  N
				34  5.20e-6  N
				35  1.25e-6  N
				36  3.16e-6  N
				37  8.64e-7  N
				38  8.64e-7  N
				41  2.68e-6  N
				43  4.82e-6  N
				44  2.99e-6  N
				45  2.99e-6  N 
				46  1.24e-6  N
				47  4.63e-6    N
				48  2.35e-6  N    	
			};
			\addlegendentry{Non-accelerated}
			
			\addplot[scatter,only marks,%
			scatter src=explicit symbolic]%
			table[meta=label] {
				x y label
				3	8.09e-7    G
				4   9.66e-7    G
				27   9.81e-6    G
				28   8.81e-6    G
				30   6.17e-6    G
				33   5.20e-6    G
				39   1.36e-5    G
				40   4.46e-6    G
				49   9.59e-6    G
			};
			\addlegendentry{GPU-accelerated}
			
			\addplot+[ mark=diamond,  draw=webgreen] table[y={create col/linear regression={y=Y}},
			meta=label,    /pgf/number format/read comma as period
			] {
				x Y label
				2	1.99e-7	A
				42  1.72e-6    A
				50  2.77e-6    A
			};
			\addlegendentry{Regression of accelerated}
			\addplot+[ mark=triangle,  draw=webgreen] table[y={create col/linear regression={y=Y}},
			meta=label,    /pgf/number format/read comma as period
			] {
				x Y label
				1	3.27e-8	N
				5  1.10e-7   N
				6  1.59e-6    N
				7  1.51e-6    N
				8  1.04e-7  N
				9  2.19e-7  N
				10  1.22e-6  N
				11  6.4e-7  N
				12  3.64e-6  N
				13  4.03e-6 N
				14  3.06e-6  N
				15  8.05e-7  N
				16  2.75e-6  N
				17  1.68e-6  N
				18  1.56e-6  N
				19  1.22e-6  N
				20  1.64e-6  N
				21  3.76e-7  N
				22  7.84e-7  N
				23  4.38e-7  N
				24  2.25e-6  N
				25  6.11e-7  N
				26  6.11e-7  N
				29  3.04e-6  N
				31  9.32e-7  N
				32  2.45e-6  N
				34  5.20e-6  N
				35  1.25e-6  N
				36  3.16e-6  N
				37  8.64e-7  N
				38  8.64e-7  N
				41  2.68e-6  N
				43  4.82e-6  N
				44  2.99e-6  N
				45  2.99e-6  N 
				46  1.24e-6  N
				47  4.63e-6    N
				48  2.35e-6  N    	
			};
			\addlegendentry{Regression of nonaccelerated}
			
			\addplot+[ mark=square,  draw=webgreen] table[y={create col/linear regression={y=Y}},
			meta=label,    /pgf/number format/read comma as period
			] {
				x Y label
				3	8.09e-7    G
				4   9.66e-7    G
				27   9.81e-6    G
				28   8.81e-6    G
				30   6.17e-6    G
				33   5.20e-6    G
				39   1.36e-5    G
				40   4.46e-6    G
				49   9.59e-6    G
			};
			\addlegendentry{Regression of GPU accelerated}
			\end{axis}
			\end{tikzpicture}
			
			\\
		\end{tabular}
	}
	\caption{Correlation of performance of processors using accelerator with separate address space and effective parallelism with ranking, in 2017.}
	\label{fig:PerformanceVSranking}
\end{figure*}

\subsection{The effect of accelerator}
\label{sec:accelerator}

As suggested by Equ.~(\ref{eq:Ppayload}), the trivial way to increase supercomputers'
resulting payload performance is to increase the single-processor
performance of its processors. Given that the single processor performance has reached its limitations,
some accelerators are frequently used for this goal. Fig.~\ref{fig:PerformanceVSranking} 
shows how using accelerators with separate address space influences ranking of supercomputers\footnote{The shared address space has two effects. On the one side, it increases the efficiency of the computing performance of the \gls{PU} by an extra factor of 3\dots4, in addition to the effect of the \gls{GPU}s with non-shared address space. On the other side, it reduces the internal latency and so reduces the efficiency dependence on the number of cores. However, not enough data exists.}.

The left subfigure shows that supercomputers' ranking does not depend on which acceleration method it uses. 
Essentially the same is confirmed by the right subfigure:
the non-payload portion raises with the ranking position, and the slope is the same for any acceleration. As the left subfigure depicts, \gls{GPU}-accelerated processors increase the payload performance of the system by a factor of 2-2.5 (the value is in good accordance with that measured in~\cite{Lee:GPUvsCPU2010})\footnote{That improved performance is about 40 times smaller than the expectation based on the nominal performance of the \gls{GPU} accelerator. The proliferation of \gls{GPU} accelerated systems made computing more power-hungry, so it also increased their power expectation, see Fig.~1.a in~\cite{BrainMasterPlan:2022}}.
The right subfigure discovers that the non-payload to payload ratio of \gls{GPU}-accelerated systems is nearly an order of magnitude worse than that of the non-accelerated systems. That is, the resulting efficiency can be (depending on the size of the system) \textit{worse} than in the case of utilizing unaccelerated processors; this can be a definite disadvantage when \gls{GPU}s used in a system with a vast number of processors. See also section~\ref{sec:accuracy}, on the one side,
how introducing \gls{GPU}s to the system increased their payload performance;
on the other side, how changing  \gls{GPU} to another type, with a higher nominal
performance, but larger memory to be filled with copied content, affected 
payload performance of the system.

\begin{figure*}
	\includegraphics[width=\textwidth]{
		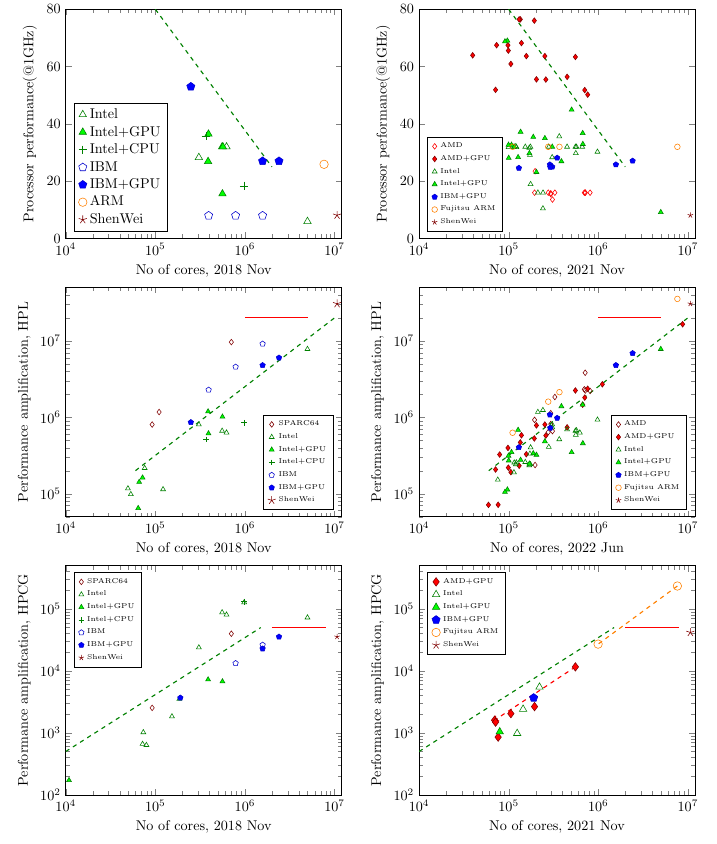}
	\caption{The correlation of the specific single processor performance (top row) and non-payload to payload efficiency (middle row: \gls{HPL}, bottom row: \gls{HPCG}) with the number of cores in the system. The empty figures 
		mark non-accelerated, and the filled figures
		GPU accelerated processor systems. The present (as of November 2021) state (right column) versus the state three years ago (left column). The red line section displays the estimated theoretically possible best non-payload to payload ratio (see section \ref{sec:estimatingfactors}). The dashed line serves as a guide for the eye.
		\label{fig:PerformanceVNoOfCores}
	}
\end{figure*}

The 'weak scaling' provides a quick and easy way to estimate the \textit{nominal} performance of a designed system. However, we experience an \textit{empirical efficiency}, see Fig.~\ref{fig:EffDependence2020Log}, when measuring  actual systems' \textit{payload} performance, which refutes the validity of 'weak scaling'.
With the proliferation of supercomputers using the same \gls{GPU}-accelerator and 
based on the same processor in systems of various sizes, we can refute the fallacy that an added \gls{GPU} accelerator multiplies the payload performance of the system's cores
by the nominal performance of the \gls{GPU}.
At the same time, we can more safely estimate the final empirical limits of supercomputing (using the present computing paradigm). 

Fig.~\ref{fig:PerformanceVNoOfCores} 
shows how the specific single-processor performance
(the single processor performance divided by the clock frequency) and the non-payload to payload ratio change in the function of the number of system's cores. The figures also show how a better integration with the \gls{GPU} components
(and the intention to achieve higher performance) transformed the landscape in three years.

As the top row displays, \textit{the accelerated specific processor performance (unlike power consumption) decreases to the level of non-accelerated specific processor performance as the systems' size increases}.
We could guess this tendency already three years ago, but now it has become much more expressed with the appearance of masses of \gls{GPU}-accelerated systems. On the one side, the enhanced acceleration technology is susceptible to the implementation method (shows a significant variance), but it can produce a performance boost factor of about 4. The dedicated measurements provided a factor 2\dots3 only~\cite{Lee:GPUvsCPU2010} but in some implementations of supercomputers, the enhancement is marginal (see the upper left subfigure of Fig.~\ref{fig:PerformanceVNoOfCores}).
On the other side, the required non-payload to payload ratio strongly decreases as the number of the cores in the system increases (also reflecting the effect of the physical size); and it is different for different workloads. 

As the middle and bottom rows display, using an accelerator seems not to influence how the amplification factor depends on the number of cores: the increased latency (caused by the need to copy data between the address spaces) counterbalances the higher processor performance at those core numbers. 
In the case of the benchmark \gls{HPL}, the dashed line (serving to guide the eye) crosses the estimated performance limit (the short red line section, see subsection~\ref{sec:estimatingfactors}) at about 10M.
Its meaning is that this could be the maximum number of cores that can cooperate \textit{in the case of this workload}. In the case of the benchmark \gls{HPCG}, the performance roofline is about 200 times lower, and the critical number of cores has a ten times lower value.
The red diamonds impressively populated the figure, but they all belong to a relatively low number (a few hundred thousand) of cores. There is no chance they can work with cores above a million. The right subfigures underpin that \textit{a high number of processors must be accompanied with good parallelization efficiency}. Otherwise, the large number of cores cannot counterbalance the decreasing efficiency, see Eq.~(\ref{eq:Ppayload}). We cannot build more extensive systems using the present paradigm and technology.

The two measured data points above the rooflines mark the supercomputers Fugaku and Taihulight. The latter uses 'cooperating processors'~\cite{CooperativeComputing2015}; both are slightly deviating from the classic computing paradigm. The former can perform better because of the cleverly placed cache memories (closer to the 'in-memory computing').
We can increase the \textit{nominal performance} without limitation. Still, the system either will not start up at all (mainly because of the "communication collapse"~\cite{CommunicationCollapse:2018})
or will work only at marginal computing and energetic efficiency. For the \gls{HPCG} workload, because of the reasons mentioned,
only reliable data are included.

Noticeably, in Fig.~\ref{fig:EffDependence2020Log} the systems having the best efficiency values 
do not use accelerator: the \textit{efficiency} of systems using accelerators is much lower also 
in the case of \gls{HPL} benchmark, but it is more disadvantageous 
in the case of \gls{HPCG} benchmark. As can be seen, one can reach the
"roofline" efficiency with using only a fragment of the available cores.
The two new items in Fig.~\ref{fig:EffDependence2020Log} (as of June 2021, Selene and Jewels; based on entirely \gls{GPU}s) not only show the worst efficacy for \gls{HPL} benchmark,
but their \gls{HPCG} efficiencies push back the 
number of usable cores (for real-life tasks) well below
the hundred thousand limit.
On the one side, the achieved \gls{HPCG} efficiency values show
the same tendency as the \gls{HPL} efficiency values: the more cores,
the lower efficiency. On the other side, figures~\ref{fig:EffDependence2020Log}~and~\ref{fig:PerformanceVNoOfCores} show
that for real-life tasks, the reasonable size of supercomputers (even in the case of benchmark \gls{HPL})
is below 1M core for non-accelerated cores.
At that number of cores, their efficiency is just a few percent, and 
\textit{increasing the number of cores decreases their efficiency}
(does NOT increase their payload performance).
In other words, we can conclude that \textit{the accelerators enable the systems to reach a much worse non-payload to payload ratio}.
For an example, see Perlmutter. The slight increase in both the nominal and payload performance for benchmark \gls{HPL} (the \gls{HPL} roofline reached) is not accompanied by any enhancement for benchmark \gls{HPCG} (the \gls{HPCG} roofline already arrived at a small fragment of cores).

In the past few years, there has appeared the tendency to attempt to show up better efficiency values when excessive supercomputers attack real-life tasks; the right bottom figure shows why.
The theoretically estimated amplification factor is about 200-300 times lower for the \gls{HPCG} workload than that
for the \gls{HPL} workload. However, some \gls{HPCG} data points seem to appear with a much better amplification factor. The reason is that the \gls{HPCG} workload does not enable to increase the system's \textit{payload performance} above its roofline when using the same number of cores that were used when benchmarking the systems with \gls{HPL} workload.
Because of this limitation, the "measured cores" in the \gls{HPCG}
benchmarking is just a fragment (about 10~\%) of the total cores. The figure shows two connected data point pairs. The upper point is the published performance (calculated assuming that the total number of cores was used) the lower point is the actual performance (it was calculated using the data "measured cores").
This correction puts back those performance amplification data to the correct scale. It confirms that for real-life tasks, the maximum number of processors in the system is below 1M for non-accelerated cores and below 0.1M cores for accelerated ones. 
Unfortunately, the item "measured cores" is not published anymore in the TOP500 lists' records. This cheating can be seen only in that the \gls{HPCG}/\gls{HPL} performance ratio is greatly improved; even in the case of older machines, there is a sudden jump (about ten times better value) in their performance ratio.

\subsection{The effect of reduced operand length}
\label{sec:operandlength}

The so-called \textit{HPL-AI} benchmark\footnote{It is a common fallacy that benchmark \textit{HPL-AI}
is benchmark for \gls{AI} systems. It means "\textit{The High Performance LINPACK for Accelerator Introspection}" (HPL-AI), and "that benchmark seeks to highlight the convergence of HPC and artificial intelligence (AI) workloads", see https://www.icl.utk.edu/hpl-ai/. It has not much to do with \gls{AI}, except that it uses the operand length common in \gls{AI} tasks. \gls{HPL}, similarly to \gls{AI}, is a \textit{workload type}. However, even 
\textit{https://www.top500.org/lists/top500/2020/06/} mismatches \textit{operand length} and \textit{workload}: "In single or further reduced precision, which are often used in machine learning and AI applications, Fugaku's peak performance is over 1,000 petaflops (1 exaflops)". 
} used Mixed Precision
rather than Double Precision computations. The name suggests that \gls{AI} applications
may run on the supercomputer with that efficiency. However, the type of workload does not change, so that one can expect the same overall behavior for \gls{AI} applications, including \gls{ANN}s, than for double-precision operands.
For \gls{AI} applications, the limitations remain the same
as described above;
except that when using Mixed Precision,
the efficiency can be better by a factor of $2\cdots 3$, strongly depending on the workload of the system; 
see also Fig.~\ref{fig:FP16vsFP64HPCG}.

Unfortunately, when using half-precision, \textit{the enhancement comes from accessing fewer data in memory and using faster operations on shorter operands, instead of reducing communication intensity\footnote{On the contrary, the relative weight of communication increased in this way.} that defines the efficiency}.
Similarly, exchanging data directly between processing units~\cite{CooperativeComputing2015} (without using global and even local memory)
also enhances $\alpha$ (and payload performance)~\cite{TaihulightHPCG:2018},
but it represents a (slightly) different computing paradigm.
Only the two mentioned measured data fall above
the limiting roofline of $(1-\alpha)$ in Figure~\ref{fig:InterconnectionVsPerformance}.

Recent supercomputers  $Fugaku$~\cite{DongarraFugakuSystem:2020} and $Summit$~\cite{MixedPrecisionHPL:2018} provided their \gls{HPL} performance 
for both 64-bit and 16-bit operands. Of course, their performance seems to be much better with a shorter operand length
(at the same number of operations, the total measurement time is much shorter).
One expected that their performance should be four times higher when using four times shorter operands.
Power consumption data~\cite{MixedPrecisionHPL:2018} underpin the expectations:
power consumption is about four times lower for four times shorter operands.
Computing performance, however, shows a slighter performance enhancement: 3.01 for $Summit$
and 3.42 for $Fugaku$ because of the needed housekeeping.

In the long run, a $Time_{X}$ value
comprises housekeeping and computation. We assume that housekeeping (indexing, branching)
is the same fixed amount for different operand lengths, and the other time contribution 
(data delivery and bit manipulation) is 
proportional with the operand length.
Given that according to our model, the measured payload performance directly reflects the sum of all contributions,
we can assume that

\begin{equation}\label{eq:measured}
\begin{alignedat}{2}
Time_{16} &= F_0 + F_{16}
\\
Time_{64} &= F_0 + 4\cdot F_{16}
\end{alignedat}
\end{equation}

\noindent where $F_0 $ is the time contribution from housekeeping (in a long-term run, using benchmark \gls{HPL}), and $F_{16}$ is 
the time contribution due to manipulating 16 bits.

We can use two simple models when calculating the relative values of $F_0$
and $F_{16}$. 
Both models are speculative rather than
technically established ones, but they nicely point to the vital point:
the role of housekeeping.

\begin{figure*}
\begin{tabular}{cc}
	\includegraphics[width=0.55\textwidth]{
		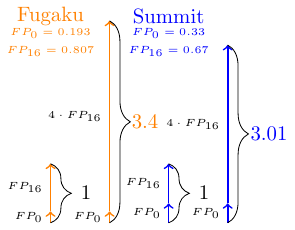}
	&
	\includegraphics[width=0.45\textwidth]{
		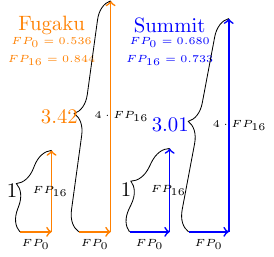}
\end{tabular}
\caption{Left: The serial summing of payload and non-payload contributions (assumes no parallelization and/or no blocking. Right: The parallel summing of payload and non-payload contributions (assumes parallelization and enables to consider mutual blocking)\label{fig:HalfPrecisionSerial}}
\end{figure*}

\begin{figure*}
\includegraphics[width=\textwidth]{
	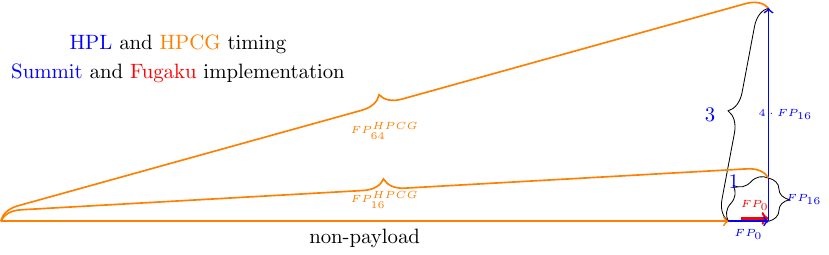}
\caption{The role of non-payload contribution in defining $HPL$ and $HPCG$ efficiencies,
	for double and half precision floating operation modes. For visibility, a hypothetic efficiency ratio $E_{HPL}/E_{HPCG}$=10 assumed. The housekeeping (including transfer time and cache misses)
	dominates, the length of operands has only marginal importance.\label{fig:FP16vsFP64HPCG}}
\end{figure*}

If no part of housekeeping runs in parallel with floating computing,
furthermore, computing and data transfer operations do not block each other,
we can use the simple summing above; see Fig.~\ref{fig:HalfPrecisionSerial}.
With the proper parallelization method, we can perform the non-floating housekeeping for the next operand in parallel with the floating operation of the current operand. 
In this way, theoretically, we can reach the possible ratio of four.
Notice, however, that this model is not aware of the data transfer time.

In the case of considering the temporal behavior of the components, we can use the trigonometric sum of the non-payload and payload times, see~\textbf{\cite{VeghTemporal:2020}}.

\begin{equation}\label{eq:measuredB}
\begin{alignedat}{2}
Time_{16} &= \sqrt{F_0^2 + F_{16}^2}
\\
Time_{64} &= \sqrt{F_0 ^2+ (4\cdot F_{16})^2}
\end{alignedat}
\end{equation}

Table~\ref{tab:EfficiencyData} shows the data calculated from the published values in the TOP500 list 
for supercomputers $Fugaku$ and $Summit$, together with their
parameters calculated as described above, using the time-aware summing model.
There is no significant difference between the values derived using the two models in the discussed simple workload case. The values in the figures are given in units of payload performance ratios; the values in the table are given in units of the respective $\alpha$ contributions. We can compare the numbers only as discussed below.

The values $Eff_{64}$ and $Eff_{16}$ are calculated from the corresponding published $R_{Max}/R_{Peak}$ values.
We calculate Amdahl's parameter using Eq.~(\ref{eq:alphafromr}) for the two different
operand lengths. As discussed, $(1-\alpha)$ is the sequential portion of the total
measurement time (aka non-payload to payload ratio). Assuming that the time unit is the total measurement time, 
the limiting time of performing a floating operation with 64-bit operands
$Time_{64}$ (on our arbitrary time scale) is directly derived from the $(1-\alpha_{64})$ value.
To get a $Time_{16}$ value on the same scale, we must correct the measured $(1-\alpha_{16})$
values for their differing measurement times (the measured
performance ratios 3.42 and 3.01, respectively).

\begin{table}
\caption{Floating point characteristics of supercomputers $Fugaku$ and $Summit$\label{tab:EfficiencyData}}
{\scriptsize
	\begin{tabular}{|p{23pt}|p{18pt}|p{18pt}|p{32pt}|p{32pt}|p{25pt}|p{25pt}|p{28pt}|p{28pt}|}
		\hline
		\hline
		\textbf{Name}   & $Eff_{64}$ & $Eff_{16}$& $(1-\alpha
		_{64})$& $(1-\alpha
		_{16})$&$Time_{64}$&$Time_{16}$&$F_{16}$&$F_{0}$\\
		\hline
		Fugaku & 0.808 & 0.691 & 3.25e-8&6.12e-8&3.25e-8&1.79e-8&\textbf{0.49e-8}&\textbf{1.3e-8}\\
		\hline
		Summit &0.74 & 0.557 & 14.7e-8&33.2e-8&14.7e-8&11.0e-8&\textbf{1.23e-8}&\textbf{9.77e-8}\\
		\hline
		\hline
	\end{tabular}
}
\end{table}

One cannot compare the absolute values of the data in the last two columns of the table directly
with those of the other supercomputer (their measurement time was different).
Given that the task and the computing model were the same, we can directly compare the ratios of values $F_{16}$ and $F_0$. The proportions of both $F_0$ values and
$F_{16}/F_0$ values show that housekeeping is much better for $Fugaku$ than for $Summit$.
Given that their architectures are globally similar, 
the plausible reason for the difference in their efficacy (and performance) is
that in the case of $Summit$, the processor core is in a role of proxy (and in this way
it represents a bottleneck), while $Fugaku$ uses "assistant cores".
The housekeeping increases latency and significantly decreases the performance of the system. 
The plausible reason for $Fugagu$s better $F_{16}$ values
is the clever use (and positioning! see~\textbf{\cite{VeghTemporal:2020}}) of L2-level cache memories.

In the case of $Summit$, we also know its \gls{HPCG} efficiency. In its case the $Time_{64}^{HPCG}$
value is $2.08\cdot e-5$, i.e. several hundred times higher than $Time_{64}^{HPL}$.
Given that $F_{16}$ and $F_{64}$ are the same in the case of the two benchmarks, the difference is caused by $F_0$.
In the long run, \textit{the different workload (iteration, more intensive communication, 
"sparse" computation forcing different cache utilization) forces different $F_0$
value, and that leads to "different efficiencies"~\cite{DifferentBenchmarks:2017}
of supercomputers under different workloads}.
Given that we used a single "snapshot" of measurement times, these values are \textit{average} values taken for all floating operations instead of actual operating times.

$Fugaku$ used
only a fraction of its cores in the benchmark \gls{HPCG},
so we can validate only the achieved payload performance (but not its efficiency). 
It is very plausible that \gls{HPCG} performance reached its roofline,
and --because of the higher number of cores-- its real \gls{HPCG} efficiency would be around that of $Taihulight$. Anyhow, it would not be fair either to assume that speculation or to accept their value measured at a different number of cores.
There is no measured data.

These data directly underpin that technology is (almost) perfect:
contribution from the benchmark calculation \textit{HPCG-FP64}
is by orders of magnitude larger\footnote{Recall here that 
cache behavior may be included} than the contribution from all the rests.
Recalling that the benchmark program imitates the behavior (as defined
by the resulting $\alpha$) of real-life programs, one can see that 
\textit{the contribution from other computing-related actors is
about a thousand times smaller than the contribution of
the computation+communication}.

The unique role of "mixed precision" efficiencies (the third kind of efficiencies of a supercomputer), see the red 'x' marks in Fig.~\ref{fig:InterconnectionVsPerformance}, deserves special attention.
Strictly speaking, we cannot correctly position the points in the figure; they belong to a different scale (they are measured on a different \gls{HW}).
On the one side, the same number of operations are performed, using the same amount of \gls{PU}s.
On the other side, four times fewer data are transferred and manipulated.
The nominal performance is expected to be four times higher than in the case of using
double-precision operands.
Without correcting for the more than three times shorter measurement (see below),
the efficiency mark is slightly above the corresponding value measured with double length operands
(the relative weight of $F_0$ is higher);
with correction, it is somewhat below it.
This difference, however, is noticeable only with benchmark \gls{HPL};
in the case of \gls{HPCG} workload, computation (including operand length) has
a marginal effect. In the former case, contributions of computation and communication
are in the same range of magnitude, competing for dominating the system's performance.  In the latter case, communication dominates performance; computing has a marginal role.
\textit{This difference is the reason why no data are available for the \gls{HPCG} benchmark
using half-precision operands.}

Fig.~\ref{fig:FP16vsFP64HPCG} illustrates the role of non-payload performance with respect to operand length.
In the figure (for visibility), a hypothetic ratio 10 of efficiencies measured by benchmarks $HPL$ and $HPCG$
was assumed. The non-payload contribution blocks the operation of the floating-point unit.
The dominating role of non-payload contribution also means that it is of little importance if we use double or half-precision operands in the computation.
The blue vectors essentially represent the case of $Summit$, the red vector represents
the $FP_0$ value of $Fugaku$ (transformed to scaling of $Summit$).
We attribute the difference between their $HPL$ performances to their different
$FP_0$ values. This difference, however, gets marginal as the workload approaches real-life conditions.

\subsection{Further efficiency values}
\label{sec:other efficiencies}

The performance corresponding to $\alpha_{HPL}^{FP0}$ is slightly above 1 EFlops (when making no floating operations, i.e., rather Eops). Another peak performance reported\footnote{https://www.olcf.ornl.gov/2018/06/08/genomics-code-exceeds-exaops-on-summit-supercomputer/} when running 
genomics code on Summit (by using a mixture of numerical precisions and mostly non-floating point instructions) is 1.88 Eops, corresponding to $\alpha_{Genom}^{FP0} = 1*10^{-8}$. Those two values refer to a different mixture of instructions,
so the agreement is more than satisfactory.

Our simple calculations result in, that \textit{in the case of the benchmark \gls{HPL}}, $FP_{0}$ values are 
in the order of $FP_{16}$, and that \textit{the benchmark \gls{HPL} is computing bound:
reducing housekeeping (including communication) has some sense for "racing supercomputers",
for real-life applications it has only marginal effect}.
On the other side, increasing the housekeeping (more communication such as in the case
of benchmark \gls{HPCG} or \gls{ANN}s) degrades the apparent performance.
At a sufficiently large amount of communication~\textbf{\cite{VeghScalingANN:2021}},
the housekeeping dominates the performance, and the contribution of $FP_X$ becomes marginal.
For large \gls{ANN} applications, using $FP_{16}$ \textit{operands} makes no real difference,
their \textit{workload} defines their performance (and efficiency).
The "commodity supercomputers" can achieve the same payload performance,
although they have a much lower number of \gls{PU}s: The "racing supercomputers"
either cannot use their impressively vast number of cores at all,
or using all their cores does not increase their payload performance in solving
real-life tasks.

\subsection{The effect of clock period's length}
\label{sec:quantaltime}

The behavior of time in computing systems
is in parallel with the quantal nature of energy, known from modern science. Time in computing systems
passes in discrete steps rather than continuously.
This difference is not noticeable under usual conditions:
both human perception and macroscopic computing operations are a million-fold longer.
Under the extreme conditions represented by many-many core systems, however, the quantal nature is the source of the inherent limitation of parallelized sequential systems. The fundamental issue is that operations must be synchronized; asynchronous operation provides performance advantages~\cite{IBMAsynchronousAPI2017}.

The need to synchronize operations (including those
of many-many processors) using a central clock signal
is especially disadvantageous when attempting to imitate the behavior of biological systems without such a central signal. Although the intention to provide asynchronous operating mode was a major design point~\cite{SpiNNaker:2013}, the hidden 
synchronization (mainly introduced by thinking in conventional \gls{SW} solutions) led to very poor efficiency~\cite{NeuralNetworkPerformance:2018},
when the system attempted to perform its
flagship goal: simulating the functionality of
(a large part of) the human brain.

As we discussed in section~\ref{sec:estimatingfactors},
the performance also depends on the length of measurement time,
because of fixed-time contributions.
When making only 10 seconds long measurements, 
the smaller denominator (compared to \gls{HPL} benchmarking time)
may result in up to $10^3$ times worse $(1-\alpha_{eff})$ and performance gain values. The dominant limiting factor, however, is a different one.

In brain simulation, a $1~ms$ integration time (essentially, a sampling time) is commonly used~\cite{NeuralNetworkPerformance:2018}. The biological time (when events happen) and
the computing time (how much computing time is required to perform
computing operations to describe the same) are not only different but also not directly related.
Working with "signals from the future" must be excluded.
For this goal, at the end of this period, one must communicate the calculated new values of the state variables to all (interested) fellow neurons.
This action essentially introduces a "biology 
clock signal period" million-fold longer than the electronic clock signal period. Correspondingly,
the achievable performance is desperately low:
the system can simulate less than $10^5$ neurons,
out of the planned $10^9$  ~\cite{NeuralNetworkPerformance:2018}\footnote{Despite its failure, the SpiNNaker2 is also under construction~\cite{SpiNNaker2:2018}}. For a detailed discussion see~\textbf{\cite{VeghBrainAmdahl:2019,VeghTemporal:2020,VeghScalingANN:2021}}.

The researchers also investigated~\cite{NeuralNetworkPerformance:2018} their power consumption efficiency. 
We presumed that --to avoid obsolete energy consumption-- they performed the measurement
at the point where involving more cores increases
the power consumption but does not increase the payload 
simulation performance.
This assumption resulted in the "reasoned guess" for the efficiency of brain simulation
in Figures~\ref{fig:EffDependence2020Log} and~\ref{fig:RooflineBrain}.
Given that using \gls{AI} workload (for a discussion from this point of view see~\textbf{\cite{VeghScalingANN:2021}}) on supercomputers is of growing 
importance, the performance gain of a \textit{processor-based} \gls{AI} application
can be estimated to be between those of \gls{HPCG} and brain simulation, closer to that of \gls{HPCG}. As discussed experimentally in~\cite{DeepNeuralNetworkTraining:2016} and theoretically in~\textbf{\cite{VeghScalingANN:2021}}, in the case of neural networks 
(especially in the case of selecting improper layering depth)
the efficiency can be much lower\footnote{ https://www.nextplatform.com/2019/10/30/cray-revamps-clusterstor-for-the-exascale-era/ :
"\textit{artificial intelligence, \dots it's the most disruptive workload from an I/O pattern perspective}"}
than that estimated value.

Recall that since \gls{AI} nodes usually perform simple calculations
compared to the functionality of supercomputer benchmarks,
their communication/calculation ratio is much higher, 
making the efficacy even worse. 
The experimental research~\cite{DeepNeuralNetworkTraining:2016}  underpins our conclusions :
\begin{itemize}
\item "strong scaling is stalling after only a few dozen nodes"
\item "The
scalability stalls when the compute times drop below the communication
times, leaving compute units idle. Hence becoming an communication bound
problem." 
\item "network layout
has a large impact on the crucial communication/compute
ratio: shallow networks with many neurons per layer \dots scale worse than deep networks with less neurons."
\end{itemize}		

The massively "bursty" nature of data (different nodes of the layer
want to use the communication subsystem simultaneously) also makes the case harder.
The commonly used global bus is overloaded with messages (for a detailed discussion see~\textbf{\cite{VeghTemporal:2020}}). That overload may lead to
a "communicational collapse" (demonstrated in Figure~5.(a) in ~\cite{CommunicationCollapse:2018}): at an extremely large number of cores, exceeding the critical threshold of communication intensity, leads to an unexpected and drastic change of network latency.

\section{Accuracy and reliability of our model}\label{sec:accuracy}

As the value of parameters of our model are inferred from non-dedicated
single-shot measurements, their reliability is limited.
One can verify, however, how our model predicts values
derived from later measurements. Supercomputers usually do not have a long lifespan and several documented stages. 
One of the rare exceptions is supercomputer \textit{Piz Daint}, and its documented lifetime spans over six years. Different amounts of cores, without and with acceleration, using various accelerators, were used at that period.

\begin{figure}
	\maxsizebox{\textwidth}{!}
	{
		\begin{tabular}{cc}
			\includegraphics[scale=1.1]{
				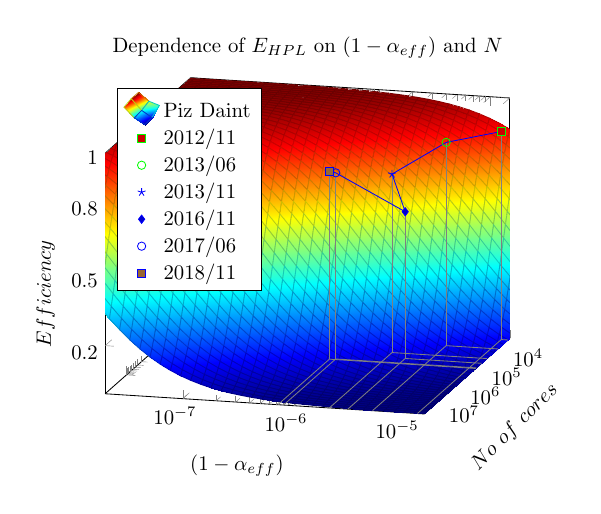}
			&
			\includegraphics[scale=.785]{
				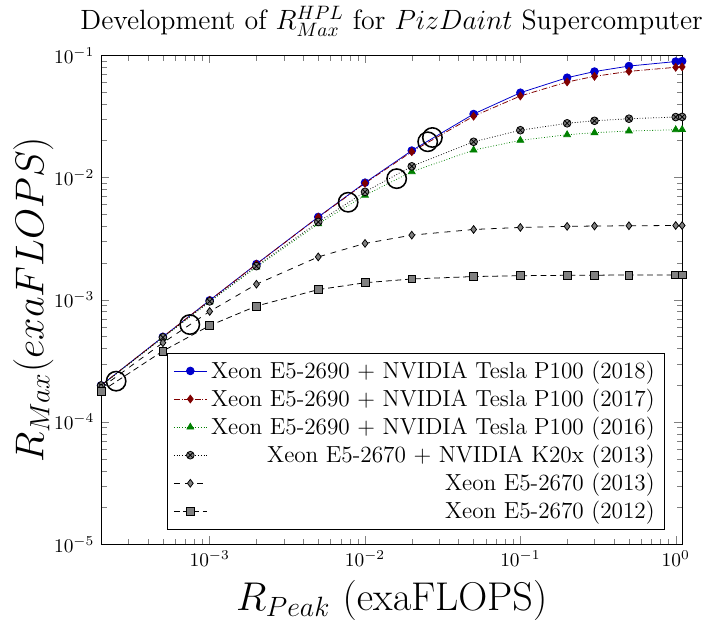}
		\end{tabular}
	}
	\caption{History of supercomputer Piz Daint in terms of efficiency and payload performance~\cite{TOP500}. 
		The left subfigure shows how the efficiency changed
		as developers proceeded towards higher performance. The right subfigure shows the reported 
		performance data (the bubbles), together with 
		diagram line calculated from the value as described above. Compare the value of diagram line
		to measured performance data in the next reported stage. 
		\label{fig:PizDaintEfficiency}
	}	
\end{figure}

Figure~\ref{fig:PizDaintEfficiency} depicts the performance and efficiency values
published during its lifetime, together with diagram lines predicting (at the time of making the prediction) values
at higher nominal performance values.
The left subfigure shows how changes made in the configuration affected its efficiency (the timeline starts in the top right corner, and a line connects adjacent stages). 

In the right subfigure, bubbles represent 
data published in adjacent editions of the TOP500 lists,
the diagram lines crossing them are predictions made from that snapshot.
We shall compare the predicted value to the value
published in the next list.
It is especially remarkable that 
introducing GPGPU acceleration resulted only in a slight increase (in good agreement with~\cite{Lee:GPUvsCPU2010}
and~\cite{WhyNotExascale:2014})
compared to the value expected based purely on the increase in the number of cores. Although between our "samplings", more than one parameter changed,
that is, the net effect cannot be demonstrated clearly, the
measured data sufficiently underpin our limited validity conclusions and show that theory correctly describes
the tendency of development of performance and efficiency, and even its predicted performance values are 
reasonably accurate.

Introducing GPU accelerator is a one-time performance increasing step~\cite{WhyNotExascale:2014}, and the theory cannot take it into account.
Notice that introducing an accelerator increased the
payload performance but decreased efficiency (copying data from one address space to another increases latency). 
Changing the accelerator to another type with slightly higher performance (but higher latency due to its larger GPGPU memory) caused a slight \textit{decrease} in the absolute performance because of the considerably dropped efficiency.

\section{Towards zettaflops}\label{sec:zettaflops}
\begin{figure}
	\includegraphics[width=\textwidth]{
		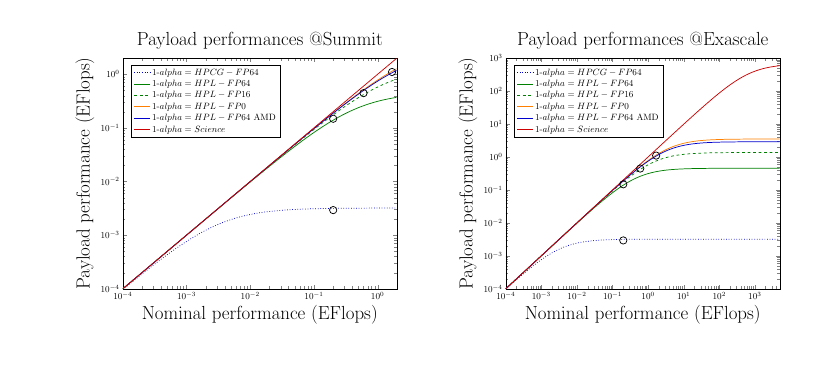}
	
	\caption{The tendency of development of payload performance in the near and farther future of supercomputing. The parameters of $Summit$ 
		are used for illustration, and for comparison, the double-precision performance diagram line of $Frontier$ is also displayed.
		The diagram lines are calculated from the theory, the
		marks are the measured values from the database TOP500~\cite{TOP500} and~\cite{MixedPrecisionHPL:2018}. \label{fig:SummitFuture}
	}
\end{figure}

As detailed above, our theoretical model enables us to calculate the
payload performance in \textit{first-order approximation} at any nominal performance value. 
In the light of all of this, one can estimate a short time and a long term development of supercomputer performance, see Figs.~\ref{fig:SummitFuture}.A and \ref{fig:SummitFuture}.B.
We calculated the diagram lines using Eq.~(\ref{eq:soverk}), with $\alpha$ parameter values derived from TOP500 data of Summit supercomputer; the bubbles show measured values.
The diagram lines from the bottom up show the double floating precision \gls{HPCG}, \gls{HPL} and the half precision~\cite{MixedPrecisionHPL:2018} \gls{HPL} (HPL-AI) diagrams.
Given that we calculated our parameter values  from a snapshot and that the calculation is essentially
an extrapolation, furthermore that at high nominal performance values using the second-order approximation is more and more pressing,
the predictions shown in Figure~\ref{fig:SummitFuture}
are rough and very optimistic approximations. However, they show values somewhat similar to actual upper limiting values.

In addition to the measured and published performance data,
two more diagram lines representing two more calculated $\alpha$ values are also depicted.
The 'FP0' (orange) diagram line is calculated with the assumption that the supercomputer makes the stuff needed to perform the \gls{HPL} benchmark,
but actual FP operations are not performed. In other words, the computer works with zero-bit length floating operands (FP0)\footnote{The role of $\alpha_{HPL}^{FP0}$ is akin to the execution time of the "empty loop" in programming.}.

The 'Science' (red) diagram line is calculated with the assumption that nothing is calculated, but science (the finite propagation time due to the limited speed of light) limits the payload performance\footnote{100 m cable length was assumed, which means $10^6$ processors pro cm and some GW dissipation.}.
The nonlinearity of the payload performance around the Eflops nominal performance is visible and depends on the amount of computing+communication and nominal performance (represented by the number of cores).

Figure~\ref{fig:SummitFuture}B shows the farther future (in first-order approximation): towards Zflops~\cite{ChinaExascale:2018}. No surprise that all payload performance diagram lines run into saturation, even the 'FP0' and 'Science' ones. 
For comparison, double-precision payload performance of $Frontier$ is also displayed: the higher single-processor performance and lower internal latency did not change the general behavior of supercomputing.
Recall that the diagram lines are calculated in the first-order approximation.
In the second-order approximation, it is expected that diagram lines reach their inflection point and break down. These top supercomputers are near to that point; this is why their development stalled.

\section{Analogies with the case of modern vs classic science}
\label{sec:Analogies}

The analogies in this section do not want to imply direct correspondence between certain physical and computing phenomena. Instead,  \textit{we wish to draw the attention to both that under extreme conditions we may encounter qualitatively different behavior, and that scrutinizing certain, formerly unnoticed or neglected aspects enables us to explain the new phenomena}.
Unlike in nature, in computing, we can change the technical implementation of critical points, and through this, we can alter the behavior of computing systems.
For a detailed discussion and more details see~\textbf{\cite{VeghModernParadigm:2019}}.

\subsection{Adding payload performances}
Eq.~(\ref{eq:Ppayload}) tells that (in first-order approximation) the speedup of a parallelized computing system
cannot exceed $1/(1-\alpha)$; a well-known consequence of Amdahl's statement.
Due to this, \textit{the computing performance of a system cannot be increased above the performance defined by the single-processor performance, the parallelization technology, and the number of processors}. \textit{The laws of nature prohibit from exceeding a specific computing performance (using the classical paradigm and its classical implementation).} \textit{There is an analogy between adding speeds in physics and adding performances in computing.} In both cases, a correction term is introduced that provides a noticeable effect only at tremendous values.
It seems to be an interesting parallel, that both nature and extremely cutting-edge
technical (computing) systems show up some extraordinary behavior,
i.e., \textit{the linear behavior experienced under normal conditions
	gets strongly non-linear at large values of the independent variable}.
That behavior makes \textit{the validity of linear extrapolations 
	as well as the linear addition of performances at least questionable} at high performance values in computing.

\subsection{Quantal nature of computing time}

In computing, the cooperation of components needs
some kind of synchronization. Typically, it uses a central clock
(with clock domains and distribution trees).
In today's technology, the length of the clock cycle is in the $ns$ range, so they seem to be quasi-continuous 
for the human perception.
In some applications, for example, when attempting to imitate
the parallel operation of the nervous system, the non-continuous nature of computing time comes to light.
In that case, the independently running neurons must be
put back to their biological scale, i.e., they stop at the
end of the "integration grid time" values, and distribute
their result to the peer neurons.
This synchronization introduces a "biological clock time", about a million-fold longer than the "processor clock time",
and limits the achievable parallelization gain.
The effect is noticeable only at a vast number of cores.
The limit is conceptual but made much worse with the
"communication burst": the appearance of the
simultaneous need for communication in the neural networks represents. For details see~\cite{VeghBrainAmdahl:2019}.

\subsection{'Quantum states' of supercomputers}\label{sec:QuantumStates}

As discussed in~\textbf{\cite{VeghModernParadigm:2019}},
the behavior of computing systems under extreme conditions
shows surprising parallels with the behavior of natural objects.
Really, "More Is Different"~\cite{MoreIsDifferent1972}.
The behavior of supercomputers
is somewhat analogous to the behavior of
quantum systems, where the measurement selects one of its possible states (and at the same time kills all other possible states). In computing, \textit{a supercomputer --as a general-purpose computing system-- has the potential of high performance defined by the impressive parameters of its components}, for all possible workflows. However, when we run a computation (that is, we measure the computing performance of our computer), that \textit{workload selects the best possible combination of limitations that
	defines its performance for the given workload and kills all other potential performances}. 

The logical dependence of the operation of the components implicitly
also means their temporal dependence~\textbf{\cite{VeghTemporal:2020}}, and it introduces idle times to computing. In this way, the workload defines how much of those potential abilities can be used: the datasheet values represent hard limits, and their workload sets the soft limits, given that 
the components block each other's operation.
\textit{The workload defines a fill-out factor:
	it introduces different idle times into component's operation, and in this way, forces workload-defined soft performance limits to the components of the supercomputer}.
Different workloads force different limitations (use their available resources differently),
giving a natural explanation of their "different efficiencies"~\cite{DifferentBenchmarks:2017}.
In other words, \textit{running some computation destroys the potentially achievable high performance, defined individually by its components}.

Benchmarking such computing systems introduces one more limiting component: the needed computation.
For floating-point computations, the 'best possible' (producing the highest figures of merit) benchmark is \gls{HPL}. With the development of parallelization and processor technology,
floating computation itself became the major contributor in defining the system's efficiency and performance.
Since the benchmark measurement method itself is a computation, 
the best measurable floating payload performance value cannot be smaller than the benchmark procedure itself represents.

For real-life programs (such as \gls{HPCG})
their workload-defined performance level (saturation value) are already set well below the Eflops nominal performance, see Fig.~\ref{fig:EffDependence2020Log}. Further enhancements in technology, such as tensor processors and OpenCAPI connection bus, can slightly increase their saturation level
but cannot change the science-defined shape of the diagram line. 

\section{Conclusion}\label{sec:Conclusions}

Supercomputers 
reached their technical limitations, their development is out of steam. To continue enhancing the components of a supercomputer
that wants to run any calculation without changing its underlying computing paradigm is not worth any more.
To enter the "next level", really renewing the classic computing paradigm is needed~\textbf{\cite{RenewingComputingVegh:2018,IntroducingEMPA2018,VeghSPAEMPA:2020,VeghTemporal:2020}}.

The ironic remark that \textit{'Perhaps supercomputers should just be required to have written in small letters at the bottom on their shiny cabinets: “Object manipulations in this supercomputer run slower than they appear.”}~\cite{DifferentBenchmarks:2017}' is becoming increasingly relevant. The impressive numbers about the performance of their components (including single-processor and/or GPU performance and speed of interconnection)
are becoming less relevant when going to the extremes.
Given that the most substantial $\alpha$ contribution today takes its origin in the
computation the supercomputer runs, even the best possible benchmark \gls{HPL} dominates floating performance measurement.
Enhancing other contributions, such as interconnection,
result in marginal enhancement of their payload performance, i.e., the
overwhelming majority of expenses increase their "dark performance" only. Because of this, the answers to the questions in the title are: \textit{there are as many performance values
	as many measurement methods (that vary with how big portion of available cores
	are used in the measurement)}, and actually \textit{benchmarks
	measure mainly how much mathematics/communication the 
	benchmark program does, rather than the supercomputer architecture}
(provided that all components deliver their technically achievable 
best parameters). 

%
%




\end{document}